\documentclass[12pt]{article}
\pdfoutput=1

\usepackage[toc,page]{appendix}
\usepackage{slashed}
\usepackage[sumlimits,intlimits,namelimits]{amsmath} 
\usepackage[english]{babel}
\usepackage{graphicx}
\usepackage{hyperref}
\usepackage{cite}

\newcommand{\ba}{\begin{eqnarray}}
\newcommand{\ea}{\end{eqnarray}}

\newcommand{\ice}[1]{\relax}

\usepackage{color}
\usepackage{amsmath}
\usepackage{booktabs} 
\usepackage{subfigure}

\def\Li{\mbox{Li}}
\begin{document}

\begin{titlepage}
  
\begin{flushright}
SI-HEP-2024-17\\
SFB-257-P3H-24-053\\
PSI-PR-24-17
\end{flushright}
\vspace{0.12cm}
\begin{center}
	  { \Large\bf
	  QCD corrections at subleading power for\\ [2mm] inclusive nonleptonic $b\rightarrow c\bar u d$ decays
}
\end{center}
\vspace{0.35cm}
\begin{center}
  {\sc Thomas Mannel${}^1$}, {\sc Daniel Moreno${}^2$},
  and {\sc Alexei A. Pivovarov${}^1$} \\[0.2cm]
  {\sf ${}^1$Center for Particle Physics Siegen,
    Theoretische Physik 1, Universit\"at Siegen\\ 57068 Siegen, Germany} \\[1mm] 
  {\sf ${}^2$Laboratory for Particle Physics, PSI Center for Neutron and Muon Sciences,
Forschungsstrasse 111, 5232 Villigen PSI, Switzerland}
\end{center}

\vspace{0.48cm}
\begin{abstract}\noindent
In this paper we compute the NLO QCD corrections for
the inclusive nonleptonic decay rates of $B$-hadrons at subleading power 
of the heavy quark expansion. In particular, we discuss the coefficient 
of the chromomagnetic moment parameter $\mu_G$ which has a large uncertainty 
at leading order due to its dependence on the renormalization scale. We find that
the renormalization-scale dependence of this coefficient is significantly reduced,
thereby reducing the theoretical uncertainty of the coefficient and consequently of the rates. The sign of the coefficient becomes stable, and we estimate that the NLO corrections will 
increase the decay rates by around $2\%$.
\end{abstract}
\end{titlepage}

\section{Introduction} 
\label{sec:Intro} 
The Heavy Quark Expansion (HQE) for processes involving bottom quarks has proven to be a tool
that allows us to perform precision calculations for processes with bottom-flavored hadrons
\cite{Chay:1990da,Bigi:1992su,Bigi:1993fe,Blok:1993va,Manohar:1993qn}. Based on the operator 
product expansion (OPE) of QCD, the HQE describes $b$-physics observables as a double-series expansion
in terms of $\alpha_s (m_b)$ and $\Lambda_{\rm QCD} / m_b$, both of which are small parameters. 
While the expansion in $\alpha_s (m_b)$ is the usual expansion in perturbative QCD, the 
expansion in $\Lambda_{\rm QCD} / m_b$ involves nonperturbative quantities, the HQE parameters, 
which are expressed in terms of forward matrix elements of local operators.

This methodology has been applied very successfully to inclusive semileptonic decays. 
Here, the leading power coefficient is known at next-to-next-to-next-to-leading
order (N$^3$LO)~\cite{Nir:1989rm,vanRitbergen:1999gs,Pak:2008qt,Pak:2008cp,Fael:2020tow,Czakon:2021ybq,Chen:2023dsi,Fael:2023tcv,Egner:2023kxw} and at next-to-next-to-leading order (N$^2$LO)~\cite{Ho-kim:1983klw,Czarnecki:1994bn,Jezabek:1996db,Jezabek:1997rk,Biswas:2009rb} in the case of a massless and massive lepton in the final state, respectively. The coefficients of the $1/m_b^2$ corrections are known at NLO~\cite{Balk:1993sz,Koyrakh:1993pq,Alberti:2013kxa,Mannel:2014xza,Mannel:2015jka,Moreno:2022goo}, and
the coefficients of the $1/m_b^3$ corrections are known at NLO~\cite{Gremm:1996df,Lenz:2013aua,Beneke:2002rj,Mannel:2019qel,Lenz:2020oce,Colangelo:2020vhu,Rahimi:2022vlv,Mannel:2021zzr,Moreno:2022goo,Moreno:2024bgq} as well. Finally, the coefficients of the $1/m_b^4$ and $1/m_b^5$ corrections are known at LO~\cite{Dassinger:2006md,Mannel:2010wj,Mannel:2023yqf}.
Based on these calculations, the theoretical uncertainy of the determination of $V_{cb}$ from 
inclusive semileptonic decays is now at the level of one percent.  

However, the theoretical status of the HQE for nonleptonic decays is less advanced. In this case, 
the leading power coefficient has been computed only recently at
NNLO~\cite{Altarelli:1980fi,Buchalla:1992gc,Bagan:1994zd,Czarnecki:2005vr,Krinner:2013cja,Egner:2024azu}. The coefficients of the
$1/m_b^2$ corrections are known at LO~\cite{Bigi:1992su,Blok:1992hw,Blok:1992he} and at NLO in the case of massless quarks in the final state~\cite{Mannel:2023zei}. The coefficients of the $1/m_b^3$ corrections are known at LO for the two-quark
operators~\cite{Lenz:2020oce,Mannel:2020fts,Moreno:2020rmk}
and at NLO for the four-quark operators~\cite{Beneke:2002rj,Franco:2002fc}. Finally, the coefficients of the $1/m_b^4$
corrections are known at LO for the four-quark operators~\cite{Gabbiani:2004tp}.

In general, the perturbative series expansion in QCD may contain special contributions
related to the infrared renormalon structure of Feynman diagrams~\cite{tHooft:1977xjm}.
The contribution of this subset can be sizable and can compromise an explicit
convergence of perturbation theory in some cases. Their resummation in all orders
of the $\alpha_s$ expansion could  be useful and can be technically
performed for particular observables~\cite{Beneke:1998ui,Krasnikov:1996jq}.
At NLO and even at NNLO however one does not expect that the perturbative
coefficients are saturated by the infrared renormalon contributions.

The inclusive nonleptonic width is an important ingredient for the calculation of heavy-hadron lifetimes in the framework of HQE. In particular, the lifetimes for bottom hadrons have been measured very precisely. The current status is~\cite{HFLAV:2022esi}
\begin{equation}
	\frac{\tau(B_s)}{\tau(B_d)}\bigg|^{{\mbox{\scriptsize exp}}} = 0.998 \pm 0.005\,,
	\;\;\;\;
	\frac{\tau(B^+)}{\tau(B_d)}\bigg|^{{\scriptsize\mbox{exp}}} = 1.076 \pm 0.004\,,
	\;\;\;\;
	\frac{\tau(\Lambda_b)}{\tau(B_d)}\bigg|^{{\scriptsize\mbox{exp}}} = 0.969 \pm 0.006\,,
\end{equation}
which calls for calculating higher orders in the HQE for inclusive nonleptonic processes to
keep up with the experimental precision.  

In this paper we compute another piece of the inclusive nonleptonic rate, building on our previous 
work~\cite{Mannel:2023zei}. We present the calculation of the  $\alpha_s (m_b)$ corrections to the perturbative coefficients of the power suppressed $\Lambda_{\rm QCD}^2/m_b^2$ terms for the total nonleptonic rate, including one massive quark in the final state. Phenomenologically this covers the quark transitions
$b \to c \bar{u} d$ and $b \to c  \bar{u} s$ and corresponds to the nonleptonic semi-inclusive 
rate for $B \to X_c$ which in principle could be measured.
This calculation could be extended as well to the case $b \to u \bar{c} s$
and $b \to u \bar{c} d$, corresponding to the 
semi-inclusive rate for $B \to X_{\bar{c}}$, but this 
is heavily CKM suppressed by $|V_{ub}|^2/|V_{cb}|^2$ and therefore we do not consider them in the present work.

All master integrals needed in the calculation
can be computed analytically in terms of polylogarithms, and thus we present a fully analytical
result. However, the expressions turn out to be involved, so we
provide a {\tt Mathematica} file \textit{``coefbcud.nb''} containing analytical results for coefficients of the nonleptonic decay rate up to order $\alpha_s \Lambda_{\rm QCD}^2/m_Q^2$ for the decay channel $b\rightarrow c \bar u d$, which are the same that for the decay channel $b\rightarrow c\bar u s$. We also provide analytical results for the master integrals needed in the calculation.

The paper is structured as follows.
In section~\ref{sec:LeffEW} we discuss the effective electroweak Lagrangian and the choice of the
renormalization scheme, including the scheme used for $\gamma_5$ and the choice of evanescent operators. Section~\ref{sec:hqerate} outlines the definitions for the HQE.
In section~\ref{sec:outline} we describe the computation. Finally, we present the results
and discuss their implications in section~\ref{sec:res}.

\section{The effective electroweak Lagrangian}
\label{sec:LeffEW}
We will consider the case in which the bottom quark decays into three quarks with different flavors,
i.e. $b \to q_1 \bar{q}_2 q_3$ with $q_1 = c$, $q_2 = u$ and $q_3 = d$ or $s$, and we shall neglect 
the masses of $q_2$ and $q_3$. In this case there are no contributions form QCD or electroweak
penguins, and thus the effective weak Lagrangian simplifies to
\begin{equation}
 \mathcal{L}_{{\scriptsize\mbox{eff}}} = - 2\sqrt{2} G_F V_{q_2 q_3} V_{q_1 b}^* (C_1 \mathcal{O}_1 + C_2 \mathcal{O}_2)
 + \mbox{h.c}\,,
 \label{eq:FermiLagr}
\end{equation}
where $G_F$ is the Fermi constant, $V_{qq'}$ are the corresponding matrix elements of the CKM matrix and $C_{1,2}$ are the Wilson 
coefficients obtained from matching (\ref{eq:FermiLagr}) to the full Standard Model at the scale $M_W$. The standard operator basis $\mathcal{O}_{1,2}$ contains color singlet and color rearranged operators~\cite{Buras:1989xd}
\begin{eqnarray}
\label{basis}
 \mathcal{O}_1 &=& (\bar b^i \Gamma_\mu q_1^j)(\bar q_2^j \Gamma^\mu q_3^i)\,,
 \\
 \mathcal{O}_2 &=& (\bar b^i \Gamma_\mu q_1^i)(\bar q_2^j \Gamma^\mu q_3^j)\,,
\end{eqnarray}
where $\Gamma_\mu=\gamma_\mu (1-\gamma_5)/2=\gamma_\mu P_L$, $(i,j)$ are color indices, and $q_{1,2,3}$ are the final state quarks. In our calculation, we will keep the 
charm quark mass $m_c$ and treat it to be of order $m_b$.

For the practical calculation, and in particular for renormalization, it is convenient to choose the operator basis
\begin{eqnarray}
 {\cal L}_{\rm eff} = -2\sqrt{2}G_F V_{q_2 q_3} V_{q_1 b}^* ( C_{+} \mathcal{O}_{+} + C_{-} \mathcal{O}_{-})
 + \mbox{h.c}\,,
 \label{LeffEWpm}
\end{eqnarray}
with 
\begin{equation}
\mathcal{O}_\pm = \frac{1}{2} (\mathcal{O}_2 \pm \mathcal{O}_1)  \quad \mbox{and} \quad  C_\pm = C_2 \pm C_1 \, .
\end{equation}
This has the advantage that this basis is diagonal under renormalization. In the $\overline{\mbox{MS}}$ renormalization scheme
\begin{eqnarray}
 C_{\pm\,,B} = Z_{\pm} C_\pm \,,\quad\quad
 Z_{\pm} = 1 - \frac{3}{N_c} (1 \mp N_c) \frac{\alpha_s(\mu)}{4\pi}\frac{1}{\epsilon}\,,
\end{eqnarray}
where the subindex $B$ stands for bare quantities and those without subscript stand for renormalized ones,
$Z_{\pm}$ are the renormalization constants of the operators $\mathcal{O}_\pm$ and $N_c=3$ is the number of colors.

The renormalization of the operators $\mathcal{O}_{\pm}$ requires to specify the treatment of $\gamma_5$ in dimensional
regularization~\cite{Altarelli:1980fi,Buchalla:1992gc,Bagan:1994zd}. In order to get a result which does not depend on the $\gamma_5$-scheme used, one has to use the same scheme for the calculation of correlators as for the
calculation of the Wilson coefficients $C_{\pm}$.

We follow the approach used by~\cite{Bagan:1994zd}, which was also used latter in our previous work for the calculation of power corrections to nonleptonic decays into massless quarks~\cite{Mannel:2023zei}. In this approach one chooses to work in standard dimensional regularization with anticommuting $\gamma_5$, also called Naive Dimensional Regularization (NDR),
using a set of evanescent
operators that preserves Fierz symmetry in the $D$ dimensional
space-time~\cite{Buras:1989xd,Dugan:1990df,Herrlich:1994kh,Jamin:1994sv,Grozin:2018wtg,Grozin:2017uto,Grozin:2016uqy}. This choice allows to circumvent the algebraic inconsistencies that arise when using anticommuting $\gamma_5$ in $D$ dimensions within closed fermion loops by avoiding the appearance of such loops.

The Wilson coefficients $C_{\pm}$ with next-to-leading logarithmic (NLL) precision
in NDR within the scheme of evanescent
operators that preserves Fierz symmetry are given by~\cite{Altarelli:1980fi,Buras:1989xd,Bagan:1994zd}
\begin{eqnarray}
	C_{\pm}(\mu) = L_{\pm}(\mu)\bigg[1 + \frac{\alpha_s(M_W) - \alpha_s(\mu)}{4\pi}R_{\pm} + \frac{\alpha_s(\mu)}{4\pi}B_{\pm}\bigg]\,,
	\label{CpmNLO}
\end{eqnarray}
which have been calculated at the scale $\mu = M_W$ and then evolved down to scales $\mu \ll M_W$ via renormalization group analysis. The equation above separates the coefficients into a scheme-independent component proportional to $R_{\pm}$
and a scheme dependent component proportional to $B_{\pm}$, with~\cite{Altarelli:1980fi,Buras:1989xd,Buchalla:1992gc,Bagan:1994zd}
\begin{eqnarray}
	R_{+} &=& \frac{10863 - 1278n_f + 80n_f^2}{6(33-2n_f)^2}\,,\quad\quad\quad\quad
	R_{-} = - \frac{15021 - 1530n_f + 80n_f^2}{3(33-2n_f)^2}\,,
	\nonumber
	\\
	B_{\pm} &=& -\frac{1}{2N_c} B (1 \mp N_c) \,, \quad\quad\quad\quad
	L_{\pm}(\mu) = \bigg(\frac{\alpha_s(M_W)}{\alpha_s(\mu)}\bigg)^{-\frac{3}{\beta_0 N_c}(1 \mp N_c)}\,,
	\label{RpmBpm}
\end{eqnarray}
where $n_f$ is the number of light flavors, $B=11$ in NDR with preservation of Fierz symmetry~\cite{Buras:1989xd} and $L_{\pm}$ is the solution of the RGE for $C_{\pm}$ to leading logarithmic (LL) accuracy with
$\beta_0 = \frac{11}{3}N_c - \frac{2}{3} n_f$.
The scheme-dependence of $B_{\pm}$ eventually cancels with the scheme-dependence of the correlators.

\section{HQE for nonleptonic decays of heavy hadrons}
\label{sec:hqerate}
By using unitarity the inclusive nonleptonic decay width $\Gamma$ of a $B$-hadron can be obtained from
the imaginary part of the forward matrix element of the transition operator ${\cal T}$
\begin{equation}\label{eq:trans_operator}
	\Gamma (B \to X) = \frac{1}{M_B} \text{Im } \langle B(p_B)|{\cal T} |B(p_B) \rangle \,,\quad\quad
	{\cal T} = i\, \int d^4 x\, T\{ {\cal L}_{\rm eff} (x)  {\cal L}_{\rm eff} (0) \} \, ,
\end{equation} 
where $p_B$ is the $B$-hadron momentum, $M_B$ its mass and $|B\rangle$ its full QCD state.

Since $m_b\gg\Lambda_{\rm QCD}$ the equation above still contains
perturbatively calculable contributions that can be separated from the non-perturbative ones
by matching the transition operator ${\cal T}$ in QCD to an expansion in $\Lambda_{\rm QCD}/m_b$ by
using Heavy Quark Effective Theory (HQET)~\cite{Mannel:1991mc,Manohar:1997qy,Georgi:1990um,Neubert:1993mb},
obtaining in this way the so-called HQE for the decay width.

To set up the HQE we closely follow our previous works and write the decay width expanded up to order
$\Lambda_{\rm QCD}^2/m_b^2$ as~\cite{Mannel:2021zzr,Moreno:2022goo,Mannel:2023zei}
\begin{equation}
	\label{hqewidth}
	\Gamma(B\rightarrow X) = \Gamma_0 |V_{q_2 q_3}|^2 |V_{q_1 b}|^2
	\bigg(  C_0
	- C_{\mu_\pi}\frac{\mu_\pi^2}{2m_b^2}
	+ C_{\mu_G}\frac{\mu_G^2}{2m_b^2} + \cdots
	\bigg)\,,
\end{equation}
where
\begin{equation}
 \Gamma_0= \frac{G_F^2 m_b^5}{192 \pi^3}\,,
\end{equation}
and the ellipsis denote terms of higher order in the $\Lambda_{\rm QCD}/m_b$ expansion.

The matching coefficients $C_i$ ($i=0,\,\mu_\pi,\,\mu_G$) can be computed as a perturbative expansion in $\alpha_s(\mu)$. We will keep the mass of the charm quark 
$m_c$, which we will treat as a quantity of the order of $m_b$. Thus the coefficients $C_i$ will depend on 
the ratio $\rho = m_c^2/m_b^2$ as well as on $\ln(\mu/m_b)$, where $\mu$ is the matching scale.

At the order we are working on we need HQET operators up to dimension five
\begin{eqnarray}
	\mathcal{O}_0 &=& \bar h_v h_v\,,\quad\quad\quad\quad\quad
	\mathcal{O}_v = \bar h_v v\cdot \pi h_v\,,
	\nonumber
	\\
    \mathcal{O}_\pi &=& \bar h_v \pi_\perp^2 h_v\,,\quad\quad\quad\quad
	\mathcal{O}_G = \frac{1}{2}\bar h_v [\gamma^\mu, \gamma^\nu] \pi_{\perp\,\mu}\pi_{\perp\,\nu}  h_v\,,\quad\quad\quad\quad
\end{eqnarray}
where $h_v$ is the HQET field, $\pi_\mu = i D_\mu = i\partial_\mu +g_s A_\mu^a T^a$ is the
QCD covariant derivative and $\pi^\mu_\perp = \pi^\mu -v^\mu (v\cdot\pi)$.

The non-perturbative HQE parameters needed to this order are $\mu_\pi^2$ and $\mu_G^2$.
These correspond to the following forward matrix elements of the local HQET operators written above~\cite{Mannel:2018mqv}
\begin{eqnarray}
	\langle B(p_B)\lvert \bar b \slashed v b \lvert B(p_B)\rangle &=& 2M_B\,,  \\
	- \langle B(p_B)\lvert \mathcal{O}_\pi \lvert B(p_B)\rangle &=& 2M_B \mu_\pi^2\,, \\
	c_F(\mu)\langle B(p_B)\lvert \mathcal{O}_G \lvert B(p_B)\rangle
	&=& 2M_B \mu_G^2\,.
\end{eqnarray}
Note also that we have included the chromomagnetic operator coefficient of the HQET Lagrangian
\begin{eqnarray}
	c_F(\mu) &=& 1 + \frac{\alpha_s(\mu)}{2\pi}\bigg[ \frac{N_c^2-1}{2N_c} + N_c\bigg(1 + \ln\bigg(\frac{\mu}{m_b}\bigg)\bigg) \bigg]
\end{eqnarray}
in the definition of $\mu_G^2$, 
which makes this HQE parameter independent of the renormalization scale $\mu$ and relates it directly 
to the mass difference of the ground-state $B$ mesons
(see e.~g. \cite{Mannel:2023zei})
\begin{equation}
\mu_G^2 = \frac{3}{4} (M_{B^*}^2 - M_B^2)\,.
\end{equation}

\section{Outline of the calculation}
\label{sec:outline}

The master expression for the decay width presented in Eq.~(\ref{hqewidth}) is derived by matching the
QCD expression for the transition operator to HQET
\begin{eqnarray}
	\label{eq:HQE-1}
	\mbox{Im } \mathcal{T} = \Gamma_0 |V_{q_2 q_3}|^2 |V_{q_1 b}|^2
	\bigg( C_0 \mathcal{O}_0
	+ C_v \frac{\mathcal{O}_v}{m_b}
	+ C_\pi \frac{\mathcal{O}_\pi}{2m_b^2}
	+ C_G \frac{\mathcal{O}_G}{2m_b^2} + \cdots
	\bigg)\,,
\end{eqnarray}
where once more the coefficients $C_i$ ($i=0,\,v,\,\pi,\,G$) depend on $\rho$ and $\ln(\mu/m_b)$ and they can be calculated as a perturbative expansion in $\alpha_s(\mu)$.

Next we trade the leading power operator $\mathcal{O}_0$
in Eq.~(\ref{eq:HQE-1}) by the local QCD operator $\bar b \slashed v b$, which is convenient
since the forward hadronic matrix element of the latter is normalized to unity and shifts the appearance of the first nonperturbative parameters to order $\Lambda_{\rm QCD}^2/m_b^2$. To that end, we need to match the QCD current to HQET
\begin{equation}
	\bar b \slashed v b = \mathcal{O}_0 + \tilde{C}_v \frac{\mathcal{O}_v}{m_b} + \tilde C_\pi \frac{\mathcal{O}_\pi}{2m_b^2}
	+ \tilde C_G \frac{\mathcal{O}_G}{2m_b^2} + \cdots  \,,
	\label{hqebvb}
\end{equation}
where ellipsis stand for higher orders in the $\Lambda_{\rm QCD}/m_b$ expansion and $\tilde{C}_i$ are the corresponding matching coefficients which depend only on $\ln(\mu/m_b)$.

Finally, we use the equation of motion of the HQET Lagrangian to eliminate the operator $\mathcal{O}_v$ in Eq.~(\ref{eq:HQE-1})
\begin{eqnarray}
	\mathcal{O}_v =
	- \frac{1}{2m_b} (\mathcal{O}_\pi+ c_F(\mu)\mathcal{O}_G) + \ldots\,,
	\label{LHQET}
\end{eqnarray}
obtaining the following expression for the inclusive decay width
\begin{eqnarray}
	\label{hqedifwidth}
	\Gamma(B \to X)
	&=& \Gamma_0 |V_{q_2 q_3}|^2 |V_{q_1 b}|^2
	\bigg[ C_0 \bigg( 1
	- \frac{\bar{C}_\pi - \bar{C}_v }{C_0}\frac{\mu_\pi^2}{2m_b^2}\bigg)
	+ \bigg(\frac{\bar{C}_G}{c_F(\mu)} -  \bar{C}_v \bigg)\frac{\mu_G^2}{2m_b^2} + \cdots \bigg] \, ,
\end{eqnarray}
where we have defined $\bar{C}_i\equiv C_i - C_0 \tilde{C}_i$ ($i=v,\,\pi,\,G$). By comparing to Eq.~(\ref{hqewidth})
we identify
\begin{eqnarray}
C_{\mu_\pi} &=& \bar{C}_\pi - \bar{C}_v\,,\quad\quad\quad\quad
C_{\mu_G} = \frac{\bar{C}_G}{c_F(\mu)} -  \bar{C}_v\,,
\end{eqnarray}
with $C_{\mu_\pi} = C_0$ implied by reparametrization invariance.

For the computation of the matching coefficients we strictly follow our previous
works~\cite{Mannel:2021zzr,Moreno:2022goo,Mannel:2023zei} where
we take the Feynman diagram amplitude of a correlator from which
the coefficient can be extracted, expand to the needed order in the small momentum $k$, and project to
the corresponding HQET operator. In that process, we use
NDR with $D=4-2\epsilon$ space-time dimensions and chose a renormalization scheme with evanescent operators preserving Fierz
symmetry up to NLO in $\alpha_s$. By making this choice one can use Fierz symmetry to write all diagrams as a
single open fermionic line without $\gamma_5$ problem. However, the matching coefficients of the HQE become scheme-dependent, but
this scheme dependence cancels against the scheme dependence of the coefficients of the effective weak Lagrangian.
For the computation we employ Feynman gauge and we use the background field method to compute the scattering amplitude
in the external gluonic field.

\begin{figure}[!htb]
	\centering
	\includegraphics[width=1.0\textwidth]{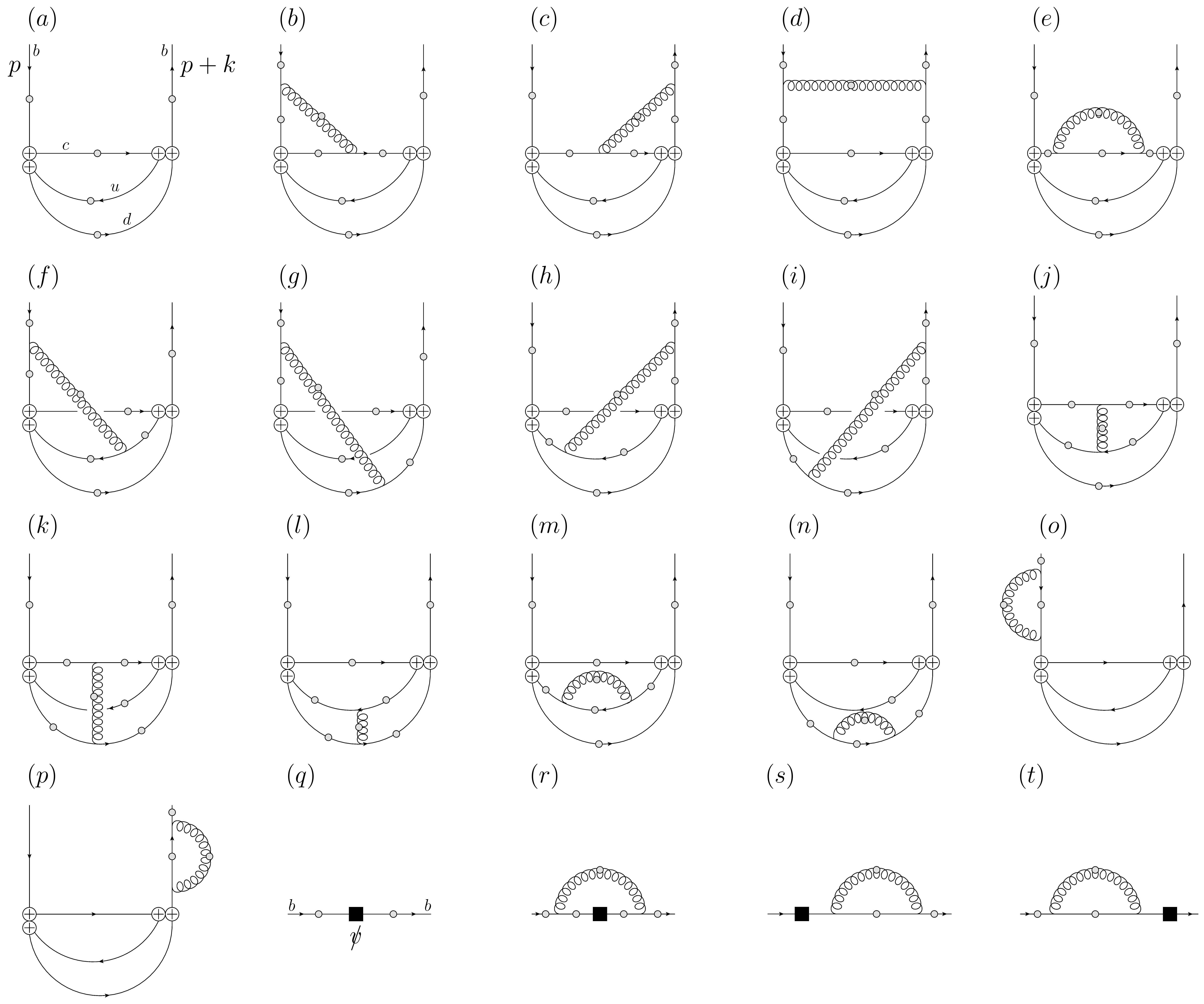}
      \caption{Feynman diagrams (a)-(t) contributing to the coefficients $C_0$, $\bar C_v$ and $\bar C_G$ of the HQE of the nonleptonic decay width up to NLO. The incoming heavy quark carries momentum $p$, with $p^2=m_b^2$. Grey dots stands for possible gluon insertions with incoming momentum $k\sim \Lambda_{\rm QCD}$. The black box vertex stands for $\slashed v$ insertions.
      The diagrams contributing to power corrections are obtained after taking into account all possible one-gluon insertions.
        Four-fermion vertices correspond $\mathcal{O}_\pm$ insertions of $\mathcal{L}_{\rm eff}$ after applying an appropriate Fierz transformation.}
                \label{NLPdiagrams}
\end{figure}

The diagrams contributing to the coefficients $C_0$, $\bar C_v$ and $\bar C_G$ are shown in Fig.~[\ref{NLPdiagrams}].
At LO and NLO in $\alpha_s$ two-loop and three-loop diagrams need to be considered, respectively.
The coefficients are obtained from the imaginary part of the corresponding diagrams.

The diagrams contributing to the leading power coefficient $C_0$ are the $b(p)\rightarrow b(p)$ two-point functions
shown in ~Fig.~[\ref{NLPdiagrams}] (a-p) by omitting gluon insertions. The diagrams contributing to the coefficients of the power corrections $\bar C_v$ and $\bar C_G$ are the $b(p)\rightarrow b(p+k)g(-k)$ three-point functions shown in ~Fig.~[\ref{NLPdiagrams}] (a-t). For the computation of $\bar C_v$ it is enough to set the gluon momentum $k=0$, whereas for $\bar C_G$ one needs to expand to
linear order in $k$. The incoming bottom quark momentum $p$ satisfies the on shell condition $p^2=m_b^2$.

In total, there are 14 diagrams contributing to $C_0$ up to NLO with 1 at LO and 13 at NLO.
For $\bar C_v$ and $\bar C_G$, there are 128 diagrams contributing up to NLO with 7 at LO and 121 at NLO.
At NLO we identify four different integral families depending on
whether the gluon exchange is between massive propagators, between a massive propagator of mass $m_b$ or $m_c$ and a massless propagator, or between massless propagators. The four integral families correspond to
diagrams (b)-(e), (f)-(i), (j)-(k) and (l)-(n), respectively.

We use integration-by-parts reduction (IBP) to write the amplitude as a combination
of the master integrals given in appendix~\ref{app:mas}. This step is automated by using {\tt LiteRed}~\cite{Lee:2012cn,Lee:2013mka}.
The LO diagram in Fig.~\ref{NLPdiagrams}~(a) can be reduced to a combination of the two 2-loop master integrals in
Fig.~\ref{masters}~(1),(2). The NLO diagrams in Fig.~\ref{NLPdiagrams}~(b)-(n) can be reduced to a combination of the 14
3-loop master integrals in Figs.~\ref{masters}~(3)-(16). The NLO diagrams in Fig.~\ref{NLPdiagrams}~(o)-(p) can also be reduced to a combination of of the two 2-loop master integrals in Fig.~\ref{masters}~(1),(2) times a 1-loop integral.
Finally diagrams (q)-(t) involve at most a 1-loop integral.

For the calculation we use several tools that allow for a high degree of automation. For the color algebra
we use {\tt ColorMath}~\cite{Sjodahl:2012nk}, whereas for Lorentz and Dirac algebra we use {\tt Tracer}~\cite{Jamin:1991dp}.
The $\epsilon$-expansion of Hypergeometric functions is achieved by using {\tt HypExp}~\cite{Huber:2005yg,Huber:2007dx}.

Finally, we make use of the $\overline{\mbox{MS}}$ renormalization scheme for $\alpha_s$ and the HQET Lagrangian, whereas
the bottom and charm quarks are renormalized on-shell. The required one-loop renormalization constants can be taken e.~g.
from~\cite{Moreno:2022goo}. Therefore, we will present our results in terms of pole masses.

\section{Discussion of the results}
\label{sec:res}

The explicit expressions for the HQE coefficients $C_0$, $\bar C_v$ and $C_{\mu_G}$ at NLO in $\alpha_s$ are rather
lengthy and therefore we provide them in the Appendix and in the supplemental material file \textit{``coefbcud.nb''}. The results for the $b\rightarrow c\bar u d$ decay channel are explicitly given in Appendix \ref{app:coefbcud}. The same results are applicable to the
$b\rightarrow c\bar u s$ decay channel.

Let us emphasize once more that, in the expressions obtained for the coefficients of the HQE, the functions multiplying the $C_{\pm}$ (or likewise the $C_{1,2}$) structures are in general dependent on the scheme used for $\gamma_5$ and the choice of the evanescent operators. This scheme dependence cancels against the scheme dependence of $C_{\pm}$ (or likewise $C_{1,2}$). Therefore, the results given in the Appendix together with the definitions given in Eqs.~(\ref{CpmNLO}) and
(\ref{RpmBpm}) provide scheme-independent coefficients.

The expression obtained for the $C_0$ coefficient at NLO agrees
with~\cite{Bagan:1994zd}, which was obtained in the same scheme we used in this work. For the power suppressed terms, our
result agrees with previous LO determinations~\cite{Bigi:1992su,Blok:1992hw,Blok:1992he}. At NLO the power suppressed terms can be
checked against our previous work~\cite{Mannel:2023zei} by taking the massless limit $m_c=0$ ($\rho=0$), and we find agreement between
both calculations. The expressions for the $\bar C_v$ and $C_{\mu_G}$ coefficients with full dependence on $m_c$ are the new
results of this paper.

Note that, in the massless case, the coefficient functions in front of the $C_1^2$ and $C_2^2$ coefficients become identical, which is implied by Fierz symmetry. This symmetry does not hold any longer once we add a massive quark,  and thus the coefficient
functions in front of $C_1^2$ and $C_2^2$ are in general different. However, for $C_0$ and $C_{\mu_G}$ at LO this symmetry is accidentally preserved and the coefficient functions in front of the $C_1^2$ and $C_2^2$ coefficients is the
same. At NLO we find that this is not true anymore.

Finally, note that the color structure of the $C_0$ and $\bar C_v$ coefficients is the same.
The reason is that, instead of considering diagrams with one-gluon insertions, the $\bar{C}_v$ coefficient can be computed by running a small momentum $k$ through the diagrams that contribute to
$C_0$ and expanding to linear order in $k$, i.~e. from the two point function $b(p+k)\rightarrow b(p+k)$.
We actually perform such a calculation and obtain the same result for $\bar C_v$, which is a strong check.

\begin{figure}[ht]
\centering
	\includegraphics[scale=0.67]{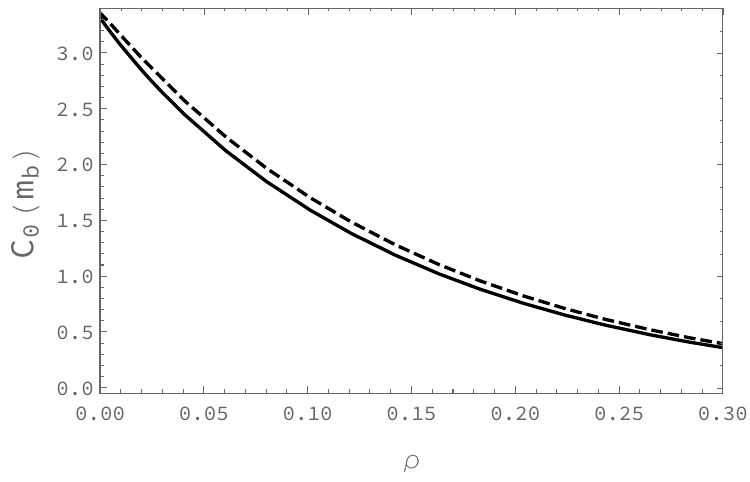}
\quad
\includegraphics[scale=0.67]{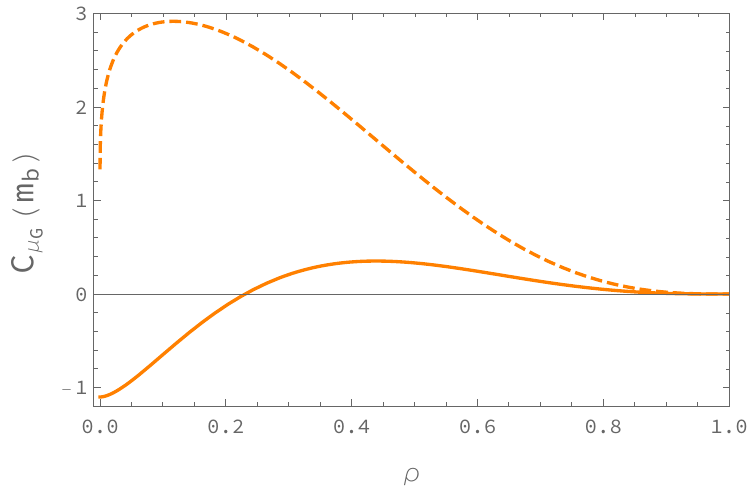}
\caption{$C_0$ (left panel) and $C_{\mu_G}$ (right panel) as a function of the mass ratio $\rho$ for $\mu=m_b$. The solid lines show the coefficients at LO precision whereas the dashed lines include NLO corrections.}
\label{fig:C0CGrho}
\end{figure}

We are now ready to discuss the impact of the $\alpha_s$ corrections and study the dependence of 
the $C_{\mu_G}$ coefficients as a
function of the renormalization scale $\mu$ and the parameter $\rho$. We will show our results in the pole scheme, in particular both masses $m_b$ and $m_c$ are taken to be pole masses. 
For the numerical evaluation~\cite{Kuhn:1998uy} we use the illustrative values
given in table~\ref{tab:par}.
\begin{table}
 \begin{center}
\begin{tabular}{|c|c|c|c|}
\hline
 Parameter & Numerical value & Parameter & Numerical value\\
   \hline
 $m_b$ & $4.7$ GeV  & $\mu_\pi^2$ & $0.5$ GeV$^2$\\
 $m_c$ & $1.6$ GeV &  $\mu_G^2$ &  $0.35$ GeV$^2$ \\
 $\rho=m_c^2/m_b^2$& $0.116$ & $M_W$ & $80.4$ GeV\\
 $\alpha_s(m_b)$ & $0.216$ & $\alpha_s(M_Z)$ & $0.120$\\
 \hline
\end{tabular}
\caption{Values of HQE parameters used in the plots of the nonleptonic rate coefficients. The
quark masses are pole masses.}
\label{tab:par}
\end{center}
\end{table}
The detailed anatomy of the coefficients in the OPE are listed in the appendix and 
can be used as a {\tt Mathematica} code supplied in the supplemental file \textit{``coefbcud.nb''}. Combining the results from the HQE with the
explicit expressions for $C_{1,2}$ (or likewise $C_{\pm}$) given in Eqs.~\ref{CpmNLO} and \ref{RpmBpm} one obtains the final expressions for
the coefficients $C_0$ and $C_{\mu_G}$. In Fig.~\ref{fig:C0CGrho} we show their dependence on the mass ratio $\rho$.
\begin{figure}[ht]
	\centering
		\includegraphics[scale=0.67]{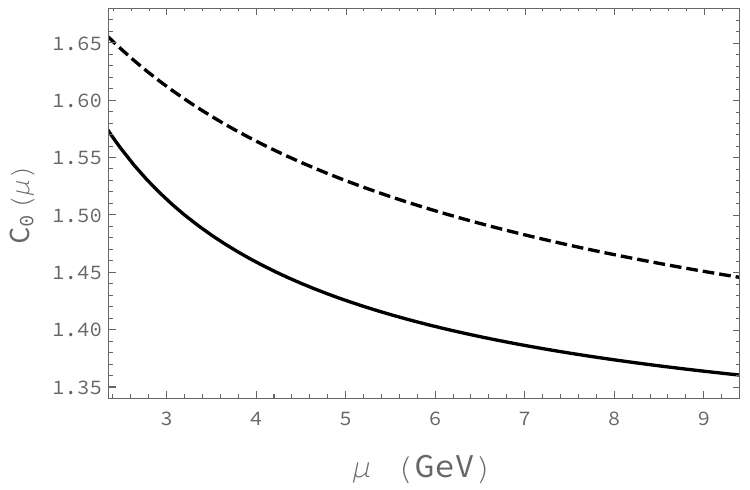}
	\quad
		\includegraphics[scale=0.67]{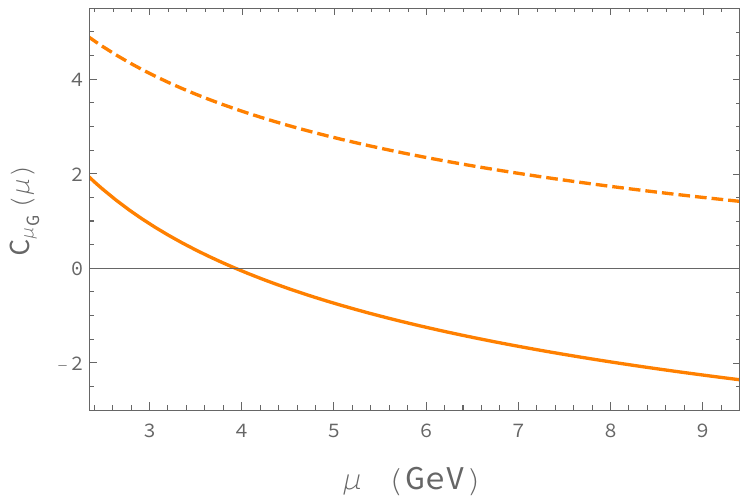}
	\caption{$C_0$ (left panel) and $C_{\mu_G}$ (right panel) as a function of the renormalization scale $\mu$ for $\rho=0.116$. The solid lines show the coefficients at LO precision and the dashed lines include NLO corrections.}
	\label{fig:C0CGmu}
\end{figure}
We find that the coefficient $C_0$ (left panel) depends quite strongly on $\rho$. Comparing to 
our previous result for $\rho = 0$ this coefficient is about a factor of two smaller, which 
originates from the reduced phase space in both the LO and NLO terms. We observe that the NLO corrections to the $C_0$ coefficient is small compared to the leading term indicating a good behaviour of the perturbative series.

The right panel shows the $\rho$ dependence of the coefficient $C_{\mu_G}$, which is very 
different for the LO and the NLO terms. In particular, for the physical value of $\rho$
and for $\mu = m_b$ (see dependence on $\mu$ below) the leading term turns out to be small
due to a cancellation of different 
contributions to this coefficient function, while this cancellation is lifted at NLO. This leads
to the fact that, at the physical value of $\rho$, the (negative) leading term is overwhelmed
by the NLO contribution which is many times larger than the LO piece and differs in sign.

In Fig.~\ref{fig:C0CGmu} we show the dependence of the coefficients $C_0$ and $C_{\mu_G}$ on the renormalization scale $\mu$ for the physical value of $\rho$. The range plotted is $m_b/2 \le \mu \le 2 m_b$ which is the typical range in which $\mu$ is varied to estimate unknown higher orders in $\alpha_s$. The coefficient $C_0$ has at LO a significant variation in the range of $\mu$ (solid line in the left panel), but we do not see a significant reduction of the $\mu$ dependence by taking into account
NLO corrections. This has been observed already previously~\cite{Bagan:1994zd}.
This has been observed already previously~\cite{Bagan:1994zd} and the reason is mainly due to the accidental absence of
dominant $\alpha_s \ln(\mu^2/M_W^2)$ terms at LL.

The $\mu$ dependence of the coefficient $C_{\mu_G}$ is more striking. Due to the cancellation mentioned already above, the leading term is small at $\mu = m_b$, but the variation in $\mu$
shows that at LO we can only conclude that $-2 \le C_{\mu_G} \le 2$ (note the very different scales 
for the y-axes in the two panels), indicating large NLO corrections. The NLO calculation reveals
that $C_{\mu_G}$ is positive and in the region $2 \le C_{\mu_G} \le 4.5$. Since the leading term suffers from cancellations and it
is thus small, the $\mu$ dependence of the NLO contribution to
$C_{\mu_G}$ remains large, even though reduced by $20\%$. We point out that this does not mean that perturbation theory in
$\alpha_s$ fails here, rather the small LO term would make a NNLO calculation necessary to
reduce the $\mu$ dependence.  

\begin{figure}[ht]
	\centering
		\includegraphics[scale=0.67]{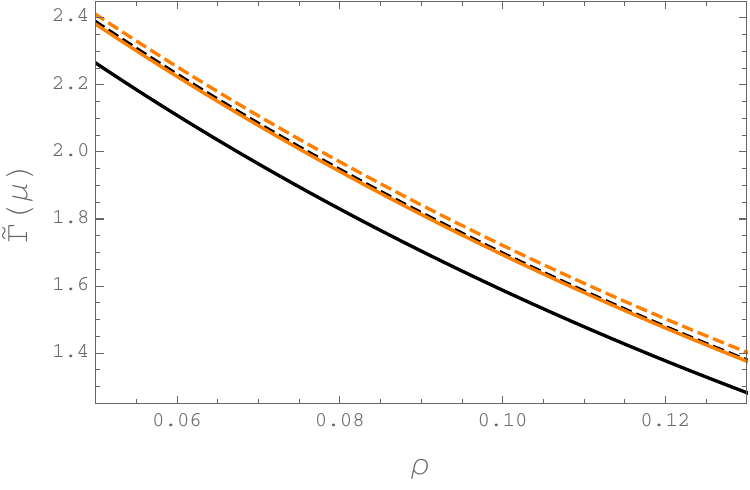}
	\quad
		\includegraphics[scale=0.67]{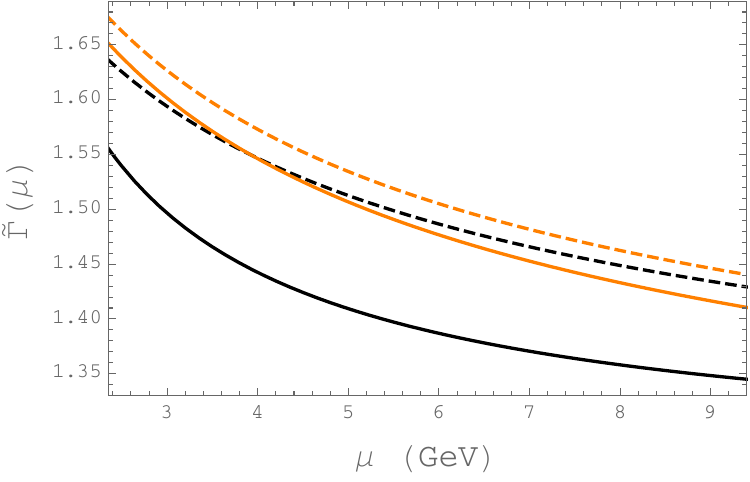}
	\caption{$\tilde{\Gamma}$ as a function of $\rho$ (left panel) for $\mu = m_b$ and as a function of $\mu$ (right panel) for the physical value of $\rho$. The black lines show the results at leading power: The solid line is the LO result and the dashed line is the NLO result. The orange lines show the results including the power corrections on top of the leading power at NLO: The solid line includes the LO result and the dashed line the NLO result.}
	\label{fig:GamT}
\end{figure}

Finally we discuss the implications for the total rate. We consider the quantity 
\begin{equation}
\tilde{\Gamma} =  \frac{\Gamma(b\rightarrow c\bar u d)}{\Gamma_0 |V_{ud}|^2 |V_{cb}|^2}
\end{equation}
and show in Fig.~\ref{fig:GamT} the dependence of $\tilde{\Gamma}$ on $\rho$ and $\mu$.  
Overall, the comparison of the dashed lines shows that the effect of the NLO contributions is small. The dependence on $\rho$ is mainly driven by the phase-space effect which is very similar to the
behaviour of the leading term. The dependence on $\mu$ is still sizable even once the NLO terms are 
included. In order to quantify the size of corrections we insert $\mu = m_b$ and the physical value of $\rho$ and
obtain with the input data from table~\ref{tab:par}
\begin{eqnarray}
\tilde{\Gamma}
 &=& C_0\bigg(1 - \frac{\mu_\pi^2}{2m_b^2}\bigg)
	+ C_{\mu_G}\frac{\mu_G^2}{2m_b^2}
  \\
 &=& (1.434_{\rm LO} + 0.105_{\rm NLO})\bigg(1 - \frac{\mu_\pi^2}{2m_b^2}\bigg)
	+ (-0.556_{\rm LO} + 3.473_{\rm NLO}) \frac{\mu_G^2}{2m_b^2}
	\\
 &=& (1.418_{\rm LO} + 0.104_{\rm NLO})_{0} + (-0.004_{\rm LO} + 0.028_{\rm NLO})_{{\mu_G}}
\end{eqnarray}
The NLO corrections to the partonic rate amount to a $7.3\%$ correction of the leading term.
The NLO corrections to the chromomagnetic operator add a $1.9\%$ correction to the leading term, 
while the 
LO contribution to the chromomagnetic operator amounts only to a $-0.3\%$ correction to the leading term, however (as argued above)
with a large uncertainty.

Considering that NNLO corrections to the partonic rate have been computed very recently and that have been found to be
about $0.3\%$ of the LO result for the decay channel under study~\cite{Egner:2024azu}, we find the NLO corrections to the chromomagnetic operator coefficient to be very important for the theoretical precision sought at present.

\section{Conclusions}
\label{sec:conc}

In this paper we have computed $\alpha_s$ corrections to the chromomagnetic operator coefficient in the HQE for the nonleptonic decay rate of $B$-hadrons. Our previous work, in which
we computed the case for massless quarks in the final state, has been extended
to the case where one quark in the final state is massive. Phenomenologically this corresponds to the contributions induced by
the quark transition $b\rightarrow c\bar u q$ and $b\rightarrow u\bar c q$ for $q = d$ or $s$. However, all decay channels rather
than $b\rightarrow c\bar u d$ are CKM suppressed and therefore we do not consider them in the current work.
Technically, all master integrals can be computed analytically and the result can be given in terms
of polylogarithms. The resulting expressions are lengthy, but we provide them as a {\tt Mathematica}
code in the ancillary file \textit{``coefbcud.nb''}.  

We have discussed the $b\rightarrow c\bar u d$ channel and find that the NLO corrections
to the coefficient of the chromomagnetic operator $C_{\mu_G}$ turn out to be very important. Since the LO contribution to $C_{\mu_G}$ suffers from a strong cancellation and a strong
dependence on the renormalization scale $\mu$, the NLO term becomes the relevant contribution to 
$C_{\mu_G}$. Although the $\mu$ dependence remains sizable, we conclude that $C_{\mu_G}$ is positive. As a phenomenological consequence, the chromomagnetic contribution will lower the (partial) lifetimes. 

In order to obtain the full NLO contributions at subleading power for nonleptonic decays, 
the next step would be to compute the case of two massive quarks in the final state, corresponding 
to e.g. the $b \to c \bar{c} s$ piece of the effective Lagrangian. From the experience with the 
present calculation this seems feasible, since it is sufficient to compute the case with the two masses being identical. 

We note that the size of the NLO corrections to subleading powers in $1/m_b$ is comparable to 
the recently calculated NNLO corrections to the partonic rate~\cite{Egner:2024azu}. To this end, 
we are getting closer to a full NLO picture for the lifetime calculations for  bottom hadrons using
the HQE. As a further perspective, we point out that the methods developed here can be also used 
to compute NLO corrections to subsubleading power, e.~g. for the coefficient of the Drawin term, which has recently been discussed intensively.

\subsection*{Acknowledgments}
This research was supported by the Deutsche Forschungsgemeinschaft
(DFG, German Research Foundation) under grant  396021762 - TRR 257
``Particle Physics Phenomenology after the Higgs Discovery''
and has received funding from the European Union’s Horizon 2020
research and innovation program under the Marie Skłodowska-Curie grant
agreement No.~884104 (PSI-FELLOW-III-3i).

\appendix

\section{Master integrals}
\label{app:mas}

\begin{figure}[!htb]
	\centering
	\includegraphics[width=0.99\textwidth]{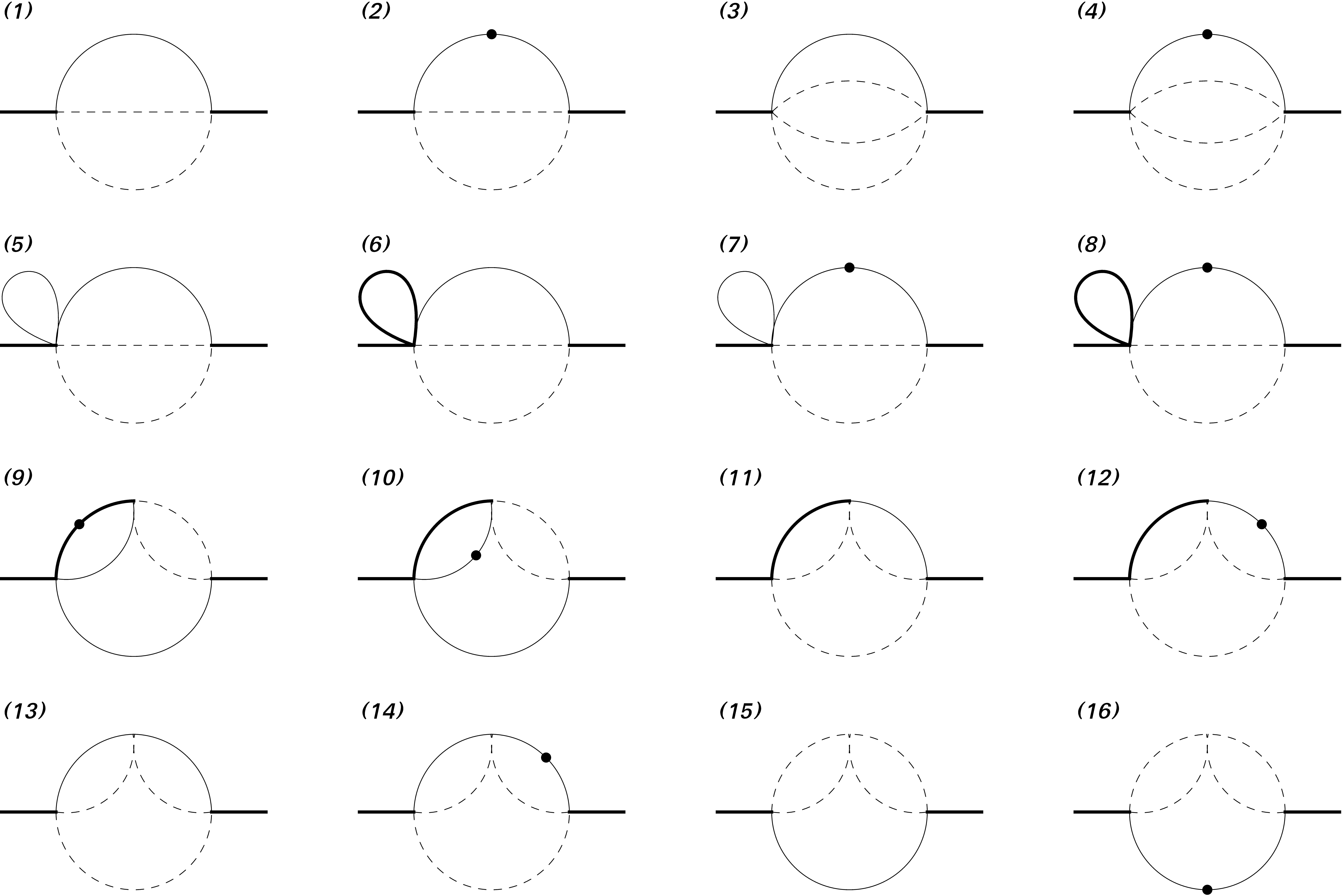}
    \caption{Master integrals (1)-(16) contributing to the matching coefficients in the HQE of the inclusive nonleptonic
    decays with a massive quark in the final state. The dashed lines are massless, the continuous thin lines have mass $m_c$ and the continuous thick lines have mass $m_b$. Black dots indicate iteration of propagators.
    The external legs are on shell, i.~e. $p^2=m_b^2$.}
    \label{masters}
\end{figure}

For completeness we provide the master integrals required for
the computation of the chromomagnetic operator coefficient of inclusive
$b \rightarrow c \bar u d$ and $b \rightarrow c \bar u s$ (also applicable to $b\rightarrow u \bar c s$ and $b\rightarrow u \bar c d$) decays expanded to the needed order in $\epsilon$. Analytical expressions are too lengthy to be displayed on the text and they can be found in the ancillary file \textit{``coefbcud.nb''}. The graphs representing the corresponding master integrals are shown
in Fig.~\ref{masters}.
After using IBP reduction we find that at LO and at NLO there are 2 two-loop and 14 three-loop self-energy-like topologies with on shell external momentum $p^2=m_b^2$, respectively.
The master integrals (1)-(2) are LO master integrals which already contribute in the semileptonic case.
The master integrals (1)-(10) also appear in the semileptonic case and they can be taken from~\cite{Mannel:2015jka}
(see also~\cite{Mannel:2021ubk}). The master integrals (11)-(16) only appear in the nonleptonic case, and therefore we compute them
explicitly in this paper.

The graphs in Fig.~\ref{masters} are defined as
\begin{equation}
 \mathcal{J}_m = \mbox{Im }
 i^{2-n} m_Q^{2n\epsilon} \bigg(\frac{e^{\gamma_E}}{4\pi}\bigg)^{n\epsilon}
  \prod_{j=1}^{n} \int\frac{d^D q_j}{(2\pi)^D}
 \prod_{k} \frac{1}{D_k}\,,\quad\quad (m=1,\ldots,16)\,,
\end{equation}
where $n$ is the number of loops and the index $j$ runs over the number of propagators in a particular graph.
The propagator signature is $D_k = q_k^2 - m_k^2$, where $q_k$ is the momentum flowing through the propagator and $m_k$ its mass.

Finally note that in the ancillary file we do not provide explicit expressions for the master integrals (9) and (10)
but for the combinations $(9)+(10)$ and $(9)-(10)$.

\section{Coefficients for $b\rightarrow c \bar u d$ channel}
\label{app:coefbcud}

We split the $b\rightarrow c \bar u d$ HQE coefficients as follows
\begin{eqnarray}
  C_i &=&
 C_1^2 C_{i,\,11}
 + C_2^2 C_{i,\,22}
 + C_1 C_2 C_{i,\,12}\,,
 \quad\quad (i=0,\,v,\,\mu_G)\,,
 \label{coefsplit1}
\end{eqnarray}
with
\begin{eqnarray}
 C_{i,\,j} &=& C_{i,\,j}^{(0)} + \frac{\alpha_s(\mu)}{\pi}C_{i,\,j}^{(1)}\,,
 \quad\quad (i=0,\,v,\,\mu_G\,;\, j=11,\,22,\,12)\,,
 \label{coefsplit2}
\end{eqnarray}
and present them in terms of the standard weak Lagrangian coefficients $C_1$ and $C_2$. For the leading power coefficient $C_0$ we
obtain
\begin{eqnarray}
C_{0,\,11}^{(0)} &=& C_{0,\,22}^{(0)} = \frac{N_c}{2}C_{0,\,12}^{(0)} = N_c (1 - 8 \rho + 8 \rho^3 - \rho^4 - 12 \rho^2 \ln\rho )\,,
\\
\nonumber
&&
\\
C_{0,\,11}^{(1)} &=&
\frac{1}{48} (N_c^2-1) \bigg(
93 - 1108 \rho + 1108 \rho^3 - 93 \rho^4
 - 8 \pi^2 (1 - 8 \rho + 8 \rho^3 - \rho^4 -12 \rho^2\ln\rho)
\nonumber
\\
&&
 - 4 (31 - 320 \rho + 320 \rho^3 - 31 \rho^4) \ln(1 - \rho)
 - 4 \rho (72 + 90\rho - 248 \rho^2 + 31 \rho^3) \ln\rho
\nonumber
\\
&&
 -24 (1 - 8 \rho - 72 \rho^2 - 8 \rho^3 + \rho^4) \ln(1 - \rho) \ln\rho
 - 12 \rho^2 (36 + 8 \rho - \rho^2) \ln^2\rho
\nonumber
\\
&&
 - 24 (1 - 8 \rho - 72 \rho^2 - 8 \rho^3 + \rho^4) \Li_2(1 - \rho)
     \bigg)\,,
\\
\nonumber
&&
\\
C_{0,\,22}^{(1)} &=&
\frac{1}{48} (N_c^2-1) \bigg(
93 - 1100 \rho + 1100 \rho^3 - 93 \rho^4
\nonumber
\\
&&
- 4 (1 - \rho^2) (17 - 64 \rho + 17 \rho^2) \ln(1 - \rho)
- 4 \rho (60 + 324 \rho - 4 \rho^2 + 17 \rho^3) \ln\rho
\nonumber
\\
&&
- 24 (1 - 12 \rho^2 + \rho^4) \ln(1 - \rho) \ln\rho
- 12 \rho^2 (36 + \rho^2) \ln^2 \rho
\nonumber
\\
&&
+ 3072 \rho^{3/2} (1 + \rho) \Li_2(1 - \sqrt{\rho})
\nonumber
\\
&&
- 24 (3 + 32 \rho^{3/2} + 48 \rho^2 + 32 \rho^{5/2} + 3 \rho^4) \Li_2(1 - \rho)
\bigg)\,,
\\
\nonumber
&&
\\
C_{0,\,12}^{(1)} &=&
 -\frac{1}{12}\frac{N_c^2-1}{2N_c} \bigg(
 153 - 1828 \rho + 1828 \rho^3 - 153 \rho^4
\nonumber
\\
&&
 + 4 (1 - \rho^2) (29 + 272 \rho + 29 \rho^2) \ln(1 - \rho)
 + 4 \rho (60 - 396 \rho + 188 \rho^2 + 47 \rho^3) \ln\rho
\nonumber
\\
&&
 + 24 (1 + 24 \rho + 60 \rho^2 + 24 \rho^3 + \rho^4) \ln(1 - \rho) \ln\rho
 + 12 \rho^2 (36 - 24 \rho + \rho^2) \ln^2\rho
\nonumber
\\
&&
 + 144 \ln\left(\frac{\mu}{m_b}\right) (1 - 8 \rho + 8 \rho^3 - \rho^4 - 12 \rho^2 \ln\rho)
 - 3072 \rho^{3/2} (1 + \rho) \Li_2(1 - \sqrt{\rho})
\nonumber
\\
&&
 + 24 (3 + 24 \rho + 32 \rho^{3/2} + 120 \rho^2 + 32 \rho^{5/2} + 24\rho^3 + 3 \rho^4) \Li_2(1 - \rho)
     \bigg)\,.
\end{eqnarray}
For the EOM operator coefficient $\bar C_v$ we obtain

\begin{eqnarray}
\bar C_{v,\,11}^{(0)} &=& \bar C_{v,\,22}^{(0)} = \frac{N_c}{2}\bar C_{v,\,12}^{(0)} =
 N_c (5 - 24 \rho + 24 \rho^2 -  8 \rho^3 + 3 \rho^4 - 12 \rho^2 \ln\rho )\,,
\\
&&
        \nonumber
        \\
\bar C_{v,\,11}^{(1)} &=&
 \frac{1}{48} (N_c^2 -1) \bigg(
  65 - 420 \rho + 1884 \rho^2 - 1772 \rho^3 + 243 \rho^4
              \nonumber
        \\
        &&
  + 24 \pi^2 (1 - 8 \rho + 8 \rho^3 - \rho^4)
  - 12 (35 - 164 \rho + 288 \rho^2 - 188 \rho^3 + 29 \rho^4) \ln(1 - \rho)
        \nonumber
        \\
        &&
       - 12 \rho (64 - 202 \rho + 120 \rho^2 - 29 \rho^3) \ln \rho
       \nonumber
        \\
        &&
       - 24 (5 - 24 \rho + 12 \rho^2 + 28 \rho^3 - 3 \rho^4) \ln(1 - \rho) \ln \rho
       - 12 \rho^2 (36 - 8 \rho + 3 \rho^2) \ln^2 \rho
        \nonumber
        \\
        &&
    - 72 (3 - 16 \rho + 8 \rho^2 + 12 \rho^3 - \rho^4 ) \Li_2(1 - \rho)
        \nonumber
        \\
        &&
       - 576 \rho^2 \Big( \ln(1 - \rho) \ln^2 \rho + \Li_2(1 - \rho) \ln \rho + 3\Li_3 (\rho) - 3\zeta(3)\Big)
    \bigg),\,
    \label{Cv22}
       \\
        \nonumber
        \\
\bar C_{v,\,22}^{(1)} &=&
 \frac{1}{48} ( N_c^2 -1) \bigg(
 65 + 604 \rho + 908 \rho^2 - 1820 \rho^3 + 243 \rho^4
        \nonumber
        \\
        &&
 - 4 (11 + 72 \rho^2 - 128 \rho^3 + 45 \rho^4) \ln(1 - \rho)
 + 4 \rho (12 + 396 \rho + 4 \rho^2 + 45 \rho^3) \ln \rho
        \nonumber
        \\
        &&
       - 24 (1 + 12 \rho^2 - 3\rho^4) \ln(1 - \rho) \ln \rho
       + 36 \rho^2 (4 + \rho^2) \ln^2 \rho
        \nonumber
        \\
        &&
    - 1536 \rho^{3/2} (1 + 3 \rho) \Li_2(1 - \sqrt{\rho})
            \nonumber
        \\
        &&
    - 24 (3 - 16 \rho^{3/2} - 48 \rho^2 - 48 \rho^{5/2} - 9 \rho^4) \Li_2(1 - \rho)
       \bigg)\,,
    \\
            \nonumber
        \\
   \bar C_{v,\,12}^{(1)} &=&
       - \frac{1}{12} \frac{N_c^2-1}{2N_c} \bigg(
       1157 - 6532 \rho + 5740 \rho^2 - 860 \rho^3 + 495 \rho^4
               \nonumber
        \\
        &&
       + 144 (5 - 24 \rho + 24 \rho^2 - 8 \rho^3 + 3 \rho^4) \ln\left(\frac{\mu}{m_b} \right)
               \nonumber
        \\
        &&
       - 4 (13 - 180 \rho + 36 \rho^2 + 212 \rho^3 - 81 \rho^4) \ln(1 - \rho)
            \nonumber
        \\
        &&
    - 4 \rho (12 + 900 \rho - 164 \rho^2 + 135 \rho^3) \ln \rho
    - 1728 \rho^2 \ln\left(\frac{\mu}{m_b} \right) \ln \rho
            \nonumber
        \\
        &&
    + 24 (1 + 48 \rho^2 - 12 \rho^3 - 3 \rho^4) \ln(1 - \rho) \ln \rho
    - 36 \rho^2 (4 - 8 \rho + \rho^2) \ln^2 \rho
            \nonumber
        \\
        &&
    +
    1536 \rho^{3/2} (1 + 3 \rho) \Li_2(1 - \sqrt{\rho})
            \nonumber
        \\
        &&
    + 24 (3 - 16 \rho^{3/2} - 12 \rho^2 - 48 \rho^{5/2} - 12 \rho^3 - 9 \rho^4) \Li_2(1 - \rho)
       \bigg)\,.
\end{eqnarray}
Finally, for the chromomagnetic operator coefficient $C_{\mu_G}$ we obtain

\begin{eqnarray}
  C_{\mu_G,\,11}^{(0)} &=& C_{\mu_G,\,22}^{(0)} =
 - N_c (3 - 8 \rho + 24 \rho^2 - 24 \rho^3 + 5 \rho^4 + 12 \rho^2 \ln\rho )\,,
 \\
 &&
 \nonumber
 \\
 C_{\mu_G,\,12}^{(0)} &=& -2 (19 - 56 \rho + 72 \rho^2 - 40 \rho^3 + 5 \rho^4 + 12 \rho^2 \ln\rho )\,,
 \\
 &&
 \nonumber
 \\
C_{\mu_G,\,11}^{(1)} &=&
     -\frac{1}{432 \rho}
  \bigg[
   \rho\Big(
 3N_c^2  (5 + 876 \rho + 10068 \rho^2 - 11948 \rho^3 + 999 \rho^4)
\nonumber
\\
&&
  - 19551 + 52516 \rho - 123160 \rho^2 + 94812 \rho^3 - 4617 \rho^4
  \Big)
\nonumber
\\
&&
 + 8\pi^2 \rho \Big(
  9N_c^2 (5 - 32 \rho + 32 \rho^2 - 5 \rho^4)
 - 109 + 576 \rho - 288 \rho^2 - 224 \rho^3 + 45 \rho^4
 \Big)
\nonumber
\\
&&
       - 20736 (1 - \rho)^3 \rho \ln\left(\frac{\mu}{m_b}\right)
\nonumber
\\
&&
    - 4\Big(
     9N_c^2 \rho (59 - 108 \rho + 1044 \rho^2 - 1076 \rho^3 + 81 \rho^4)
\nonumber
\\
&&
     - 12 - 1141 \rho + 3564 \rho^2 - 20508 \rho^3 + 18880 \rho^4 - 783 \rho^5
     \Big) \ln(1 - \rho)
\nonumber
\\
&&
    - 4 \rho^2 \Big( 9N_c^2 ( 48 - 618\rho + 672 \rho^2 - 81 \rho^3)
\nonumber
\\
&&
    - 2520 + 16470 \rho - 12760 \rho^2 + 783 \rho^3 \Big) \ln\rho
\nonumber
\\
&&
      - 24 \rho \Big(
       9N_c^2 (3 - 12 \rho + 112 \rho^2 + 56 \rho^3 - 5 \rho^4)
\nonumber
\\
&&
      +  61 + 360 \rho - 2220 \rho^2 - 924 \rho^3 + 45 \rho^4
          \Big) \ln(1 - \rho) \ln\rho
\nonumber
\\
&&
      + 12 \rho^2 \Big(
      9N_c^2 \rho (36 + 40 \rho - 5 \rho^2)
      +  72 - 516 \rho - 952 \rho^2 + 45 \rho^3
      \Big) \ln^2 \rho
\nonumber
\\
&&
       - 6144 \rho^{3/2} (3 - 10\rho) \Li_2(1 - \sqrt{\rho})
\nonumber
\\
&&
       - 24 \rho \Big(
       9N_c^2 (7 - 32 \rho + 140 \rho^2 + 44 \rho^3 - 5 \rho^4)
\nonumber
\\
&&
       +  61 - 192 \rho^{1/2} + 792 \rho + 640 \rho^{3/2} - 2412 \rho^2 - 716 \rho^3 + 45 \rho^4
          \Big) \Li_2(1 - \rho)
\nonumber
\\
&&
      - 576 \rho^3 ( 3N_c^2 (1 - \rho) - 9 + \rho)
      \nonumber
\\
&&
\times
      \Big(
        \ln(1 - \rho) \ln^2\rho + \frac{\pi^2}{6} \ln\rho + \Li_2(1 - \rho)\ln\rho + 3\Li_3(\rho) - 3\zeta(3)
      \Big)
\nonumber
\\
&&
       - 5184  \rho^3 ( N_c^2  - 1) \frac{\pi^2}{6} \ln\rho
       \bigg]\,,
\label{CG22}
\end{eqnarray}

\begin{eqnarray}
C_{\mu_G,\,22}^{(1)} &=&
          -\frac{1}{432 \rho} \bigg[
      \rho \Big(
        3 N_c^2  (61 + 528 \rho + 2716 \rho^2 - 4304 \rho^3 + 999 \rho^4)
\nonumber
\\
&&
       - 19647 + 59996 \rho - 83384 \rho^2 + 47652 \rho^3 - 4617 \rho^4
       \Big)
\nonumber
\\
&&
       - 64 \pi^2 \rho (37 - 117 \rho + 99 \rho^2 - 19 \rho^3)
       - 20736 (1 - \rho)^3 \rho \ln\left(\frac{\mu}{m_b}\right)
\nonumber
\\
&&
    + \Big(
     12 N_c^2 (14 - 7 \rho + 168 \rho^2 - 968 \rho^3 + 970 \rho^4 - 177 \rho^5)
\nonumber
\\
&&
    - 48 - 52 \rho - 7632 \rho^2 + 40656 \rho^3 - 36992 \rho^4 + 4068 \rho^5
          \Big) \ln(1 - \rho)
 \nonumber
\\
&&
    + 4\rho^2 \Big(
     3N_c^2  (24 + 920 \rho - 802 \rho^2 + 177 \rho^3)
      \nonumber
\\
&&
    + (324 - 6048 \rho + 7556 \rho^2 - 1017 \rho^3)
      \Big) \ln \rho
 \nonumber
\\
&&
      + 24\rho \Big(
       3N_c^2  (3 + 24 \rho - 68 \rho^2 - 44 \rho^3 + 15 \rho^4)
 \nonumber
\\
&&
      + (139 - 720 \rho + 1200 \rho^2 + 300 \rho^3 - 45 \rho^4)
       \Big) \ln(1 - \rho) \ln\rho
 \nonumber
\\
&&
      + 12\rho^2\Big(
       3 N_c^2  (12 + 64 \rho - 44 \rho^2 + 15 \rho^3)
      - (72 + 60 \rho + 272 \rho^2 + 45 \rho^3)
      \Big) \ln^2\rho
 \nonumber
\\
&&
          - 3072\rho^{3/2} \Big( 3N_c^2  (1 + \rho) - 2(3 - 4 \rho) \Big) \Li_2(1 - \sqrt{\rho})
 \nonumber
\\
&&
       + 24 \rho\Big( 3N_c^2 (9 + 32 \rho^{1/2} + 48 \rho + 32 \rho^{3/2} + 24 \rho^2 - 132 \rho^3 + 45 \rho^4)
 \nonumber
\\
&&
       + (313 - 192 \rho^{1/2} - 1440 \rho + 256 \rho^{3/2} + 1836 \rho^2 + 364 \rho^3 - 135 \rho^4) \Big) \Li_2(1 - \rho)
 \nonumber
\\
&&
    + 576 \rho^3 (3 + \rho) \Big(\ln(1 - \rho) \ln^2 \rho + \frac{\pi^2}{6}\ln\rho + \Li_2(1 - \rho) \ln\rho + 3\Li_3(\rho) - 3\zeta(3) \Big)
    \bigg]\,,
\nonumber
\\
&&
\end{eqnarray}
\begin{eqnarray}
 C_{\mu_G,\,12}^{(1)} &=&
 -\frac{1}{216\rho}
  \frac{1}{N_c} \bigg[
 3 \rho \Big( N_c^2 (1335 - 9040 \rho - 4516 \rho^2 + 14912 \rho^3 - 2691 \rho^4)
\nonumber
\\
&&
 +  (3153 + 1284 \rho + 22024 \rho^2 - 28612 \rho^3 + 2151 \rho^4)
 \Big)
 \nonumber
\\
&&
 + 64\pi^2 \rho \Big(
 3N_c^2  (1 - 12 \rho + 15 \rho^2 - 4 \rho^3)
 +  (37 - 117 \rho + 99 \rho^2 - 19 \rho^3)
 \Big)
 \nonumber
\\
&&
       - 1296 \rho \Big(
         N_c^2 (3 - 8 \rho + 24 \rho^2 - 24 \rho^3 + 5 \rho^4)
          \nonumber
\\
&&
       - (19 - 56 \rho + 72 \rho^2 - 40 \rho^3 + 5 \rho^4)
       \Big) \ln\left(\frac{\mu}{m_b}\right)
      \nonumber
\\
&&
      + 12\Big(
       N_c^2 (38 - 1163 \rho + 5748 \rho^2 - 4736 \rho^3 + 326 \rho^4 - 213 \rho^5)
        \nonumber
\\
&&
      - (52 - 757 \rho + 4164 \rho^2 - 196 \rho^3 - 3032 \rho^4 - 231 \rho^5)
          \Big)\ln(1 - \rho)
\nonumber
\\
&&
          - 12 \rho^2 \Big(
           N_c^2 ( 540 - 248 \rho + 530\rho^2 - 483 \rho^3)
 \nonumber
\\
&&
          -  (820 + 2864\rho - 1508 \rho^2 - 501 \rho^3)
          \Big) \ln \rho
\nonumber
\\
&&
          - 15552\rho^3 ( N_c^2 -1) \ln\left(\frac{\mu}{m_b}\right) \ln\rho
\nonumber
\\
&&
      - 24\rho\Big(
      3 N_c^2  (29 - 216 \rho - 364 \rho^2 - 40 \rho^3 - 15 \rho^4)
       \nonumber
\\
&&
      - (19 - 72 \rho - 2472 \rho^2 - 580 \rho^3 - 45 \rho^4)
          \Big) \ln(1 - \rho) \ln \rho
\nonumber
\\
&&
       + 12\rho^2 \Big(
        3 N_c^2  (12 + 16 \rho - 216 \rho^2 + 15 \rho^3)
       - (72 - 228 \rho - 1216 \rho^2 + 45 \rho^3)
          \Big) \ln^2 \rho
\nonumber
\\
&&
            - 3072 \rho^{3/2} \Big( 3N_c^2  (1 + \rho) -  2(3 - 4 \rho) \Big) \Li_2(1 - \sqrt{\rho})
\nonumber
\\
&&
      +  24\rho \Big(
       3 N_c^2 (-55 + 32 \rho^{1/2} + 324 \rho + 32 \rho^{3/2} + 408 \rho^2 - 52 \rho^3 + 45 \rho^4)
        \nonumber
\\
&&
      - (55 + 192 \rho^{1/2} - 288 \rho - 256 \rho^{3/2} + 3060 \rho^2 + 124 \rho^3 + 135 \rho^4)
      \Big) \Li_2(1 - \rho)
          \nonumber
\\
&&
- 576 \rho^3 (3 N_c^2 (2 - \rho)+3 + \rho )
 \nonumber
\\
&&
\times\Big(
 \ln(1 - \rho)  \ln^2 \rho +  \frac{\pi^2}{6}\ln\rho + \Li_2(1 - \rho)\ln\rho + 3 \Li_3(\rho) - 3\zeta(3)
\Big)
          \bigg]\,,
\end{eqnarray}
where the additional $\pi^2/6$ term in the last line of Eq.~(\ref{CG22}) comes from the contribution of $\bar C_v$ to the $C_{\mu_G}$ coefficient, in particular from the last line of Eq.~(\ref{Cv22}). This is because the last line of Eq.~(\ref{Cv22}) does not combine in the same way the penultimate line of Eq.~(\ref{CG22}) does. More precisely, the $\pi^2$ term is not present in the last line of
in Eq.~(\ref{Cv22}).

\newpage

\bibliographystyle{JHEP}
\bibliography{NLbcudpaper090824}

\providecommand{\href}[2]{#2}\begingroup\raggedright\begin{thebibliography}{10}

\bibitem{Chay:1990da}
J.~Chay, H.~Georgi and B.~Grinstein, \emph{{Lepton energy distributions in
  heavy meson decays from QCD}},
  \href{https://doi.org/10.1016/0370-2693(90)90916-T}{\emph{Phys. Lett. B}
  {\bfseries 247} (1990) 399}.

\bibitem{Bigi:1992su}
I.I.Y.~Bigi, N.G.~Uraltsev and A.I.~Vainshtein, \emph{{Nonperturbative
  corrections to inclusive beauty and charm decays: QCD versus phenomenological
  models}}, \href{https://doi.org/10.1016/0370-2693(92)90908-M}{\emph{Phys.
  Lett. B} {\bfseries 293} (1992) 430}
  [\href{https://arxiv.org/abs/hep-ph/9207214}{{\ttfamily hep-ph/9207214}}].

\bibitem{Bigi:1993fe}
I.I.Y.~Bigi, M.A.~Shifman, N.G.~Uraltsev and A.I.~Vainshtein, \emph{{QCD
  predictions for lepton spectra in inclusive heavy flavor decays}},
  \href{https://doi.org/10.1103/PhysRevLett.71.496}{\emph{Phys. Rev. Lett.}
  {\bfseries 71} (1993) 496}
  [\href{https://arxiv.org/abs/hep-ph/9304225}{{\ttfamily hep-ph/9304225}}].

\bibitem{Blok:1993va}
B.~Blok, L.~Koyrakh, M.A.~Shifman and A.I.~Vainshtein, \emph{{Differential
  distributions in semileptonic decays of the heavy flavors in QCD}},
  \href{https://doi.org/10.1103/PhysRevD.50.3572}{\emph{Phys. Rev. D}
  {\bfseries 49} (1994) 3356}
  [\href{https://arxiv.org/abs/hep-ph/9307247}{{\ttfamily hep-ph/9307247}}].

\bibitem{Manohar:1993qn}
A.V.~Manohar and M.B.~Wise, \emph{{Inclusive semileptonic B and polarized
  Lambda(b) decays from QCD}},
  \href{https://doi.org/10.1103/PhysRevD.49.1310}{\emph{Phys. Rev. D}
  {\bfseries 49} (1994) 1310}
  [\href{https://arxiv.org/abs/hep-ph/9308246}{{\ttfamily hep-ph/9308246}}].

\bibitem{Nir:1989rm}
Y.~Nir, \emph{{The Mass Ratio m(c) / m(b) in Semileptonic B Decays}},
  \href{https://doi.org/10.1016/0370-2693(89)91495-0}{\emph{Phys. Lett. B}
  {\bfseries 221} (1989) 184}.

\bibitem{vanRitbergen:1999gs}
T.~van Ritbergen, \emph{{The Second order QCD contribution to the semileptonic
  b ---\ensuremath{>} u decay rate}},
  \href{https://doi.org/10.1016/S0370-2693(99)00407-4}{\emph{Phys. Lett. B}
  {\bfseries 454} (1999) 353}
  [\href{https://arxiv.org/abs/hep-ph/9903226}{{\ttfamily hep-ph/9903226}}].

\bibitem{Pak:2008qt}
A.~Pak and A.~Czarnecki, \emph{{Mass effects in muon and semileptonic b
  ---\ensuremath{>} c decays}},
  \href{https://doi.org/10.1103/PhysRevLett.100.241807}{\emph{Phys. Rev. Lett.}
  {\bfseries 100} (2008) 241807}
  [\href{https://arxiv.org/abs/0803.0960}{{\ttfamily 0803.0960}}].

\bibitem{Pak:2008cp}
A.~Pak and A.~Czarnecki, \emph{{Heavy-to-heavy quark decays at NNLO}},
  \href{https://doi.org/10.1103/PhysRevD.78.114015}{\emph{Phys. Rev. D}
  {\bfseries 78} (2008) 114015}
  [\href{https://arxiv.org/abs/0808.3509}{{\ttfamily 0808.3509}}].

\bibitem{Fael:2020tow}
M.~Fael, K.~Sch\"onwald and M.~Steinhauser, \emph{{Third order corrections to
  the semileptonic b\textrightarrow{}c and the muon decays}},
  \href{https://doi.org/10.1103/PhysRevD.104.016003}{\emph{Phys. Rev. D}
  {\bfseries 104} (2021) 016003}
  [\href{https://arxiv.org/abs/2011.13654}{{\ttfamily 2011.13654}}].

\bibitem{Czakon:2021ybq}
M.~Czakon, A.~Czarnecki and M.~Dowling, \emph{{Three-loop corrections to the
  muon and heavy quark decay rates}},
  \href{https://doi.org/10.1103/PhysRevD.103.L111301}{\emph{Phys. Rev. D}
  {\bfseries 103} (2021) L111301}
  [\href{https://arxiv.org/abs/2104.05804}{{\ttfamily 2104.05804}}].

\bibitem{Chen:2023dsi}
L.-B.~Chen, H.T.~Li, Z.~Li, J.~Wang, Y.~Wang and Q.-f.~Wu, \emph{{Analytic
  third-order QCD corrections to top-quark and semileptonic b\textrightarrow{}u
  decays}}, \href{https://doi.org/10.1103/PhysRevD.109.L071503}{\emph{Phys.
  Rev. D} {\bfseries 109} (2024) L071503}
  [\href{https://arxiv.org/abs/2309.00762}{{\ttfamily 2309.00762}}].

\bibitem{Fael:2023tcv}
M.~Fael and J.~Usovitsch, \emph{{Third order correction to semileptonic $b\to
  u$ decay: Fermionic contributions}},
  \href{https://doi.org/10.1103/PhysRevD.108.114026}{\emph{Phys. Rev. D}
  {\bfseries 108} (2023) 114026}
  [\href{https://arxiv.org/abs/2310.03685}{{\ttfamily 2310.03685}}].

\bibitem{Egner:2023kxw}
M.~Egner, M.~Fael, K.~Sch\"onwald and M.~Steinhauser, \emph{{Revisiting
  semileptonic B meson decays at next-to-next-to-leading order}},
  \href{https://doi.org/10.1007/JHEP09(2023)112}{\emph{JHEP} {\bfseries 09}
  (2023) 112} [\href{https://arxiv.org/abs/2308.01346}{{\ttfamily
  2308.01346}}].

\bibitem{Ho-kim:1983klw}
Q.~Ho-kim and X.-Y.~Pham, \emph{{Exact One Gluon Corrections for Inclusive Weak
  Processes}}, \href{https://doi.org/10.1016/0003-4916(84)90258-6}{\emph{Annals
  Phys.} {\bfseries 155} (1984) 202}.

\bibitem{Czarnecki:1994bn}
A.~Czarnecki, M.~Jezabek and J.H.~Kuhn, \emph{{Radiative corrections to b
  ---\ensuremath{>} c tau anti-tau-neutrino}},
  \href{https://doi.org/10.1016/0370-2693(94)01681-2}{\emph{Phys. Lett. B}
  {\bfseries 346} (1995) 335}
  [\href{https://arxiv.org/abs/hep-ph/9411282}{{\ttfamily hep-ph/9411282}}].

\bibitem{Jezabek:1996db}
M.~Jezabek and L.~Motyka, \emph{{Tau lepton distributions in semileptonic B
  decays}}, \href{https://doi.org/10.1016/S0550-3213(97)00341-6}{\emph{Nucl.
  Phys. B} {\bfseries 501} (1997) 207}
  [\href{https://arxiv.org/abs/hep-ph/9701358}{{\ttfamily hep-ph/9701358}}].

\bibitem{Jezabek:1997rk}
M.~Jezabek and P.~Urban, \emph{{Polarization of tau leptons in semileptonic B
  decays}}, \href{https://doi.org/10.1016/S0550-3213(98)00189-8}{\emph{Nucl.
  Phys. B} {\bfseries 525} (1998) 350}
  [\href{https://arxiv.org/abs/hep-ph/9712440}{{\ttfamily hep-ph/9712440}}].

\bibitem{Biswas:2009rb}
S.~Biswas and K.~Melnikov, \emph{{Second order QCD corrections to inclusive
  semileptonic b ---\ensuremath{>} X(c) l anti-nu(l) decays with massless and
  massive lepton}}, \href{https://doi.org/10.1007/JHEP02(2010)089}{\emph{JHEP}
  {\bfseries 02} (2010) 089} [\href{https://arxiv.org/abs/0911.4142}{{\ttfamily
  0911.4142}}].

\bibitem{Balk:1993sz}
S.~Balk, J.G.~Korner, D.~Pirjol and K.~Schilcher, \emph{{Inclusive semileptonic
  B decays in QCD including lepton mass effects}},
  \href{https://doi.org/10.1007/BF01557233}{\emph{Z. Phys. C} {\bfseries 64}
  (1994) 37} [\href{https://arxiv.org/abs/hep-ph/9312220}{{\ttfamily
  hep-ph/9312220}}].

\bibitem{Koyrakh:1993pq}
L.~Koyrakh, \emph{{Nonperturbative corrections to the heavy lepton energy
  distribution in the inclusive decays H(b) ---\ensuremath{>} tau anti-neutrino
  X}}, \href{https://doi.org/10.1103/PhysRevD.49.3379}{\emph{Phys. Rev. D}
  {\bfseries 49} (1994) 3379}
  [\href{https://arxiv.org/abs/hep-ph/9311215}{{\ttfamily hep-ph/9311215}}].

\bibitem{Alberti:2013kxa}
A.~Alberti, P.~Gambino and S.~Nandi, \emph{{Perturbative corrections to power
  suppressed effects in semileptonic B decays}},
  \href{https://doi.org/10.1007/JHEP01(2014)147}{\emph{JHEP} {\bfseries 01}
  (2014) 147} [\href{https://arxiv.org/abs/1311.7381}{{\ttfamily 1311.7381}}].

\bibitem{Mannel:2014xza}
T.~Mannel, A.A.~Pivovarov and D.~Rosenthal, \emph{{Inclusive semileptonic B
  decays from QCD with NLO accuracy for power suppressed terms}},
  \href{https://doi.org/10.1016/j.physletb.2014.12.058}{\emph{Phys. Lett. B}
  {\bfseries 741} (2015) 290}
  [\href{https://arxiv.org/abs/1405.5072}{{\ttfamily 1405.5072}}].

\bibitem{Mannel:2015jka}
T.~Mannel, A.A.~Pivovarov and D.~Rosenthal, \emph{{Inclusive weak decays of
  heavy hadrons with power suppressed terms at NLO}},
  \href{https://doi.org/10.1103/PhysRevD.92.054025}{\emph{Phys. Rev. D}
  {\bfseries 92} (2015) 054025}
  [\href{https://arxiv.org/abs/1506.08167}{{\ttfamily 1506.08167}}].

\bibitem{Moreno:2022goo}
D.~Moreno, \emph{{NLO QCD corrections to inclusive semitauonic weak decays of
  heavy hadrons up to 1/mb3}},
  \href{https://doi.org/10.1103/PhysRevD.106.114008}{\emph{Phys. Rev. D}
  {\bfseries 106} (2022) 114008}
  [\href{https://arxiv.org/abs/2207.14245}{{\ttfamily 2207.14245}}].

\bibitem{Gremm:1996df}
M.~Gremm and A.~Kapustin, \emph{{Order 1/m(b)**3 corrections to B
  --\ensuremath{>} X(c) lepton anti-neutrino decay and their implication for
  the measurement of Lambda-bar and lambda(1)}},
  \href{https://doi.org/10.1103/PhysRevD.55.6924}{\emph{Phys. Rev. D}
  {\bfseries 55} (1997) 6924}
  [\href{https://arxiv.org/abs/hep-ph/9603448}{{\ttfamily hep-ph/9603448}}].

\bibitem{Lenz:2013aua}
A.~Lenz and T.~Rauh, \emph{{D-meson lifetimes within the heavy quark
  expansion}}, \href{https://doi.org/10.1103/PhysRevD.88.034004}{\emph{Phys.
  Rev. D} {\bfseries 88} (2013) 034004}
  [\href{https://arxiv.org/abs/1305.3588}{{\ttfamily 1305.3588}}].

\bibitem{Beneke:2002rj}
M.~Beneke, G.~Buchalla, C.~Greub, A.~Lenz and U.~Nierste, \emph{{The $B^+
  -B^0_d$ Lifetime Difference Beyond Leading Logarithms}},
  \href{https://doi.org/10.1016/S0550-3213(02)00561-8}{\emph{Nucl. Phys. B}
  {\bfseries 639} (2002) 389}
  [\href{https://arxiv.org/abs/hep-ph/0202106}{{\ttfamily hep-ph/0202106}}].

\bibitem{Mannel:2019qel}
T.~Mannel and A.A.~Pivovarov, \emph{{QCD corrections to inclusive heavy hadron
  weak decays at $\Lambda_{\rm QCD}^3 /m_Q^3$}},
  \href{https://doi.org/10.1103/PhysRevD.100.093001}{\emph{Phys. Rev. D}
  {\bfseries 100} (2019) 093001}
  [\href{https://arxiv.org/abs/1907.09187}{{\ttfamily 1907.09187}}].

\bibitem{Lenz:2020oce}
A.~Lenz, M.L.~Piscopo and A.V.~Rusov, \emph{{Contribution of the Darwin
  operator to non-leptonic decays of heavy quarks}},
  \href{https://doi.org/10.1007/JHEP12(2020)199}{\emph{JHEP} {\bfseries 12}
  (2020) 199} [\href{https://arxiv.org/abs/2004.09527}{{\ttfamily
  2004.09527}}].

\bibitem{Colangelo:2020vhu}
P.~Colangelo, F.~De~Fazio and F.~Loparco, \emph{{Inclusive semileptonic
  $\Lambda_{b}$ decays in the Standard Model and beyond}},
  \href{https://doi.org/10.1007/JHEP11(2020)032}{\emph{JHEP} {\bfseries 11}
  (2020) 032} [\href{https://arxiv.org/abs/2006.13759}{{\ttfamily
  2006.13759}}].

\bibitem{Rahimi:2022vlv}
M.~Rahimi and K.K.~Vos, \emph{{Standard Model predictions for lepton flavour
  universality ratios of inclusive semileptonic B decays}},
  \href{https://doi.org/10.1007/JHEP11(2022)007}{\emph{JHEP} {\bfseries 11}
  (2022) 007} [\href{https://arxiv.org/abs/2207.03432}{{\ttfamily
  2207.03432}}].

\bibitem{Mannel:2021zzr}
T.~Mannel, D.~Moreno and A.A.~Pivovarov, \emph{{NLO QCD corrections to
  inclusive $b \rightarrow c \ell \bar{\nu}$decay spectra up to~$1/m_Q^3$}},
  \href{https://doi.org/10.1103/PhysRevD.105.054033}{\emph{Phys. Rev. D}
  {\bfseries 105} (2022) 054033}
  [\href{https://arxiv.org/abs/2112.03875}{{\ttfamily 2112.03875}}].

\bibitem{Moreno:2024bgq}
D.~Moreno, \emph{{QCD corrections to the Darwin coefficient in inclusive
  semileptonic
  B\textrightarrow{}Xu\ensuremath{\ell}\ensuremath{\nu}\textasciimacron{}\ensuremath{\ell}
  decays}}, \href{https://doi.org/10.1103/PhysRevD.109.074030}{\emph{Phys. Rev.
  D} {\bfseries 109} (2024) 074030}
  [\href{https://arxiv.org/abs/2402.13805}{{\ttfamily 2402.13805}}].

\bibitem{Dassinger:2006md}
B.M.~Dassinger, T.~Mannel and S.~Turczyk, \emph{{Inclusive semi-leptonic B
  decays to order 1 / m(b)**4}},
  \href{https://doi.org/10.1088/1126-6708/2007/03/087}{\emph{JHEP} {\bfseries
  03} (2007) 087} [\href{https://arxiv.org/abs/hep-ph/0611168}{{\ttfamily
  hep-ph/0611168}}].

\bibitem{Mannel:2010wj}
T.~Mannel, S.~Turczyk and N.~Uraltsev, \emph{{Higher Order Power Corrections in
  Inclusive B Decays}},
  \href{https://doi.org/10.1007/JHEP11(2010)109}{\emph{JHEP} {\bfseries 11}
  (2010) 109} [\href{https://arxiv.org/abs/1009.4622}{{\ttfamily 1009.4622}}].

\bibitem{Mannel:2023yqf}
T.~Mannel, I.S.~Milutin and K.K.~Vos, \emph{{Inclusive semileptonic $ b\to
  c\ell \overline{\nu} $ decays to order $ 1/{m}_b^5 $}},
  \href{https://doi.org/10.1007/JHEP02(2024)226}{\emph{JHEP} {\bfseries 02}
  (2024) 226} [\href{https://arxiv.org/abs/2311.12002}{{\ttfamily
  2311.12002}}].

\bibitem{Altarelli:1980fi}
G.~Altarelli, G.~Curci, G.~Martinelli and S.~Petrarca, \emph{{QCD Nonleading
  Corrections to Weak Decays as an Application of Regularization by Dimensional
  Reduction}}, \href{https://doi.org/10.1016/0550-3213(81)90473-9}{\emph{Nucl.
  Phys. B} {\bfseries 187} (1981) 461}.

\bibitem{Buchalla:1992gc}
G.~Buchalla, \emph{{O (alpha-s) QCD corrections to charm quark decay in
  dimensional regularization with nonanticommuting gamma-5}},
  \href{https://doi.org/10.1016/0550-3213(93)90081-Y}{\emph{Nucl. Phys. B}
  {\bfseries 391} (1993) 501}.

\bibitem{Bagan:1994zd}
E.~Bagan, P.~Ball, V.M.~Braun and P.~Gosdzinsky, \emph{{Charm quark mass
  dependence of QCD corrections to nonleptonic inclusive B decays}},
  \href{https://doi.org/10.1016/0550-3213(94)90591-6}{\emph{Nucl. Phys. B}
  {\bfseries 432} (1994) 3}
  [\href{https://arxiv.org/abs/hep-ph/9408306}{{\ttfamily hep-ph/9408306}}].

\bibitem{Czarnecki:2005vr}
A.~Czarnecki, M.~Slusarczyk and F.V.~Tkachov, \emph{{Enhancement of the
  hadronic b quark decays}},
  \href{https://doi.org/10.1103/PhysRevLett.96.171803}{\emph{Phys. Rev. Lett.}
  {\bfseries 96} (2006) 171803}
  [\href{https://arxiv.org/abs/hep-ph/0511004}{{\ttfamily hep-ph/0511004}}].

\bibitem{Krinner:2013cja}
F.~Krinner, A.~Lenz and T.~Rauh, \emph{{The inclusive decay $b \to c\bar{c}s$
  revisited}},
  \href{https://doi.org/10.1016/j.nuclphysb.2013.07.028}{\emph{Nucl. Phys. B}
  {\bfseries 876} (2013) 31} [\href{https://arxiv.org/abs/1305.5390}{{\ttfamily
  1305.5390}}].

\bibitem{Egner:2024azu}
M.~Egner, M.~Fael, K.~Sch\"onwald and M.~Steinhauser, \emph{{Nonleptonic
  $B$-meson decays to next-to-next-to-leading order}},
  \href{https://arxiv.org/abs/2406.19456}{{\ttfamily 2406.19456}}.

\bibitem{Blok:1992hw}
B.~Blok and M.A.~Shifman, \emph{{The Rule of discarding 1/N(c) in inclusive
  weak decays. 1.}},
  \href{https://doi.org/10.1016/0550-3213(93)90504-I}{\emph{Nucl. Phys. B}
  {\bfseries 399} (1993) 441}
  [\href{https://arxiv.org/abs/hep-ph/9207236}{{\ttfamily hep-ph/9207236}}].

\bibitem{Blok:1992he}
B.~Blok and M.A.~Shifman, \emph{{The Rule of discarding 1/N(c) in inclusive
  weak decays. 2.}},
  \href{https://doi.org/10.1016/0550-3213(93)90505-J}{\emph{Nucl. Phys. B}
  {\bfseries 399} (1993) 459}
  [\href{https://arxiv.org/abs/hep-ph/9209289}{{\ttfamily hep-ph/9209289}}].

\bibitem{Mannel:2023zei}
T.~Mannel, D.~Moreno and A.A.~Pivovarov, \emph{{Heavy-quark expansion for
  lifetimes: Toward the QCD corrections to power suppressed terms}},
  \href{https://doi.org/10.1103/PhysRevD.107.114026}{\emph{Phys. Rev. D}
  {\bfseries 107} (2023) 114026}
  [\href{https://arxiv.org/abs/2304.08964}{{\ttfamily 2304.08964}}].

\bibitem{Mannel:2020fts}
T.~Mannel, D.~Moreno and A.~Pivovarov, \emph{{Heavy quark expansion for heavy
  hadron lifetimes: completing the $ 1/{m}_b^3 $ corrections}},
  \href{https://doi.org/10.1007/JHEP08(2020)089}{\emph{JHEP} {\bfseries 08}
  (2020) 089} [\href{https://arxiv.org/abs/2004.09485}{{\ttfamily
  2004.09485}}].

\bibitem{Moreno:2020rmk}
D.~Moreno, \emph{{Completing $1/m_b^3$ corrections to non-leptonic
  bottom-to-up-quark decays}},
  \href{https://doi.org/10.1007/JHEP01(2021)051}{\emph{JHEP} {\bfseries 01}
  (2021) 051} [\href{https://arxiv.org/abs/2009.08756}{{\ttfamily
  2009.08756}}].

\bibitem{Franco:2002fc}
E.~Franco, V.~Lubicz, F.~Mescia and C.~Tarantino, \emph{{Lifetime ratios of
  beauty hadrons at the next-to-leading order in QCD}},
  \href{https://doi.org/10.1016/S0550-3213(02)00262-6}{\emph{Nucl. Phys. B}
  {\bfseries 633} (2002) 212}
  [\href{https://arxiv.org/abs/hep-ph/0203089}{{\ttfamily hep-ph/0203089}}].

\bibitem{Gabbiani:2004tp}
F.~Gabbiani, A.I.~Onishchenko and A.A.~Petrov, \emph{{Spectator effects and
  lifetimes of heavy hadrons}},
  \href{https://doi.org/10.1103/PhysRevD.70.094031}{\emph{Phys. Rev. D}
  {\bfseries 70} (2004) 094031}
  [\href{https://arxiv.org/abs/hep-ph/0407004}{{\ttfamily hep-ph/0407004}}].

\bibitem{tHooft:1977xjm}
G.~'t~Hooft, \emph{{Can We Make Sense Out of Quantum Chromodynamics?}},
  {\emph{Subnucl. Ser.} {\bfseries 15} (1979) 943}.

\bibitem{Beneke:1998ui}
M.~Beneke, \emph{{Renormalons}},
  \href{https://doi.org/10.1016/S0370-1573(98)00130-6}{\emph{Phys. Rept.}
  {\bfseries 317} (1999) 1}
  [\href{https://arxiv.org/abs/hep-ph/9807443}{{\ttfamily hep-ph/9807443}}].

\bibitem{Krasnikov:1996jq}
N.V.~Krasnikov and A.A.~Pivovarov, \emph{{Renormalization schemes and
  renormalons}}, \href{https://doi.org/10.1142/S0217732396000849}{\emph{Mod.
  Phys. Lett. A} {\bfseries 11} (1996) 835}
  [\href{https://arxiv.org/abs/hep-ph/9602272}{{\ttfamily hep-ph/9602272}}].

\bibitem{HFLAV:2022esi}
{\scshape HFLAV} collaboration, \emph{{Averages of b-hadron, c-hadron, and
  \ensuremath{\tau}-lepton properties as of 2021}},
  \href{https://doi.org/10.1103/PhysRevD.107.052008}{\emph{Phys. Rev. D}
  {\bfseries 107} (2023) 052008}
  [\href{https://arxiv.org/abs/2206.07501}{{\ttfamily 2206.07501}}].

\bibitem{Buras:1989xd}
A.J.~Buras and P.H.~Weisz, \emph{{QCD Nonleading Corrections to Weak Decays in
  Dimensional Regularization and 't Hooft-Veltman Schemes}},
  \href{https://doi.org/10.1016/0550-3213(90)90223-Z}{\emph{Nucl. Phys. B}
  {\bfseries 333} (1990) 66}.

\bibitem{Dugan:1990df}
M.J.~Dugan and B.~Grinstein, \emph{{On the vanishing of evanescent operators}},
  \href{https://doi.org/10.1016/0370-2693(91)90680-O}{\emph{Phys. Lett. B}
  {\bfseries 256} (1991) 239}.

\bibitem{Herrlich:1994kh}
S.~Herrlich and U.~Nierste, \emph{{Evanescent operators, scheme dependences and
  double insertions}},
  \href{https://doi.org/10.1016/0550-3213(95)00474-7}{\emph{Nucl. Phys. B}
  {\bfseries 455} (1995) 39}
  [\href{https://arxiv.org/abs/hep-ph/9412375}{{\ttfamily hep-ph/9412375}}].

\bibitem{Jamin:1994sv}
M.~Jamin and A.~Pich, \emph{{QCD corrections to inclusive Delta S = 1,2
  transitions at the next-to-leading order}},
  \href{https://doi.org/10.1016/0550-3213(94)90171-6}{\emph{Nucl. Phys. B}
  {\bfseries 425} (1994) 15}
  [\href{https://arxiv.org/abs/hep-ph/9402363}{{\ttfamily hep-ph/9402363}}].

\bibitem{Grozin:2018wtg}
A.G.~Grozin, T.~Mannel and A.A.~Pivovarov, \emph{{$B^0$-$\bar{B}^0$ mixing:
  Matching to HQET at NNLO}},
  \href{https://doi.org/10.1103/PhysRevD.98.054020}{\emph{Phys. Rev. D}
  {\bfseries 98} (2018) 054020}
  [\href{https://arxiv.org/abs/1806.00253}{{\ttfamily 1806.00253}}].

\bibitem{Grozin:2017uto}
A.G.~Grozin, T.~Mannel and A.A.~Pivovarov, \emph{{Towards a
  Next-to-Next-to-Leading Order analysis of matching in $B^0$-$\bar{B}^0$
  mixing}}, \href{https://doi.org/10.1103/PhysRevD.96.074032}{\emph{Phys. Rev.
  D} {\bfseries 96} (2017) 074032}
  [\href{https://arxiv.org/abs/1706.05910}{{\ttfamily 1706.05910}}].

\bibitem{Grozin:2016uqy}
A.G.~Grozin, R.~Klein, T.~Mannel and A.A.~Pivovarov, \emph{{$B^0-\bar{B}^0$
  mixing at next-to-leading order}},
  \href{https://doi.org/10.1103/PhysRevD.94.034024}{\emph{Phys. Rev. D}
  {\bfseries 94} (2016) 034024}
  [\href{https://arxiv.org/abs/1606.06054}{{\ttfamily 1606.06054}}].

\bibitem{Mannel:1991mc}
T.~Mannel, W.~Roberts and Z.~Ryzak, \emph{{A Derivation of the heavy quark
  effective Lagrangian from QCD}},
  \href{https://doi.org/10.1016/0550-3213(92)90204-O}{\emph{Nucl. Phys. B}
  {\bfseries 368} (1992) 204}.

\bibitem{Manohar:1997qy}
A.V.~Manohar, \emph{{The HQET / NRQCD Lagrangian to order alpha / m-3}},
  \href{https://doi.org/10.1103/PhysRevD.56.230}{\emph{Phys. Rev. D} {\bfseries
  56} (1997) 230} [\href{https://arxiv.org/abs/hep-ph/9701294}{{\ttfamily
  hep-ph/9701294}}].

\bibitem{Georgi:1990um}
H.~Georgi, \emph{{An Effective Field Theory for Heavy Quarks at Low-energies}},
  \href{https://doi.org/10.1016/0370-2693(90)91128-X}{\emph{Phys. Lett. B}
  {\bfseries 240} (1990) 447}.

\bibitem{Neubert:1993mb}
M.~Neubert, \emph{{Heavy quark symmetry}},
  \href{https://doi.org/10.1016/0370-1573(94)90091-4}{\emph{Phys. Rept.}
  {\bfseries 245} (1994) 259}
  [\href{https://arxiv.org/abs/hep-ph/9306320}{{\ttfamily hep-ph/9306320}}].

\bibitem{Mannel:2018mqv}
T.~Mannel and K.K.~Vos, \emph{{Reparametrization Invariance and Partial
  Re-Summations of the Heavy Quark Expansion}},
  \href{https://doi.org/10.1007/JHEP06(2018)115}{\emph{JHEP} {\bfseries 06}
  (2018) 115} [\href{https://arxiv.org/abs/1802.09409}{{\ttfamily
  1802.09409}}].

\bibitem{Lee:2012cn}
R.N.~Lee, \emph{{Presenting LiteRed: a tool for the Loop InTEgrals REDuction}},
   \href{https://arxiv.org/abs/1212.2685}{{\ttfamily 1212.2685}}.

\bibitem{Lee:2013mka}
R.N.~Lee, \emph{{LiteRed 1.4: a powerful tool for reduction of multiloop
  integrals}}, \href{https://doi.org/10.1088/1742-6596/523/1/012059}{\emph{J.
  Phys. Conf. Ser.} {\bfseries 523} (2014) 012059}
  [\href{https://arxiv.org/abs/1310.1145}{{\ttfamily 1310.1145}}].

\bibitem{Sjodahl:2012nk}
M.~Sj\"odahl, \emph{{ColorMath - A package for color summed calculations in
  SU(Nc)}}, \href{https://doi.org/10.1140/epjc/s10052-013-2310-4}{\emph{Eur.
  Phys. J. C} {\bfseries 73} (2013) 2310}
  [\href{https://arxiv.org/abs/1211.2099}{{\ttfamily 1211.2099}}].

\bibitem{Jamin:1991dp}
M.~Jamin and M.E.~Lautenbacher, \emph{{TRACER: Version 1.1: A Mathematica
  package for gamma algebra in arbitrary dimensions}},
  \href{https://doi.org/10.1016/0010-4655(93)90097-V}{\emph{Comput. Phys.
  Commun.} {\bfseries 74} (1993) 265}.

\bibitem{Huber:2005yg}
T.~Huber and D.~Maitre, \emph{{HypExp: A Mathematica package for expanding
  hypergeometric functions around integer-valued parameters}},
  \href{https://doi.org/10.1016/j.cpc.2006.01.007}{\emph{Comput. Phys. Commun.}
  {\bfseries 175} (2006) 122}
  [\href{https://arxiv.org/abs/hep-ph/0507094}{{\ttfamily hep-ph/0507094}}].

\bibitem{Huber:2007dx}
T.~Huber and D.~Maitre, \emph{{HypExp 2, Expanding Hypergeometric Functions
  about Half-Integer Parameters}},
  \href{https://doi.org/10.1016/j.cpc.2007.12.008}{\emph{Comput. Phys. Commun.}
  {\bfseries 178} (2008) 755}
  [\href{https://arxiv.org/abs/0708.2443}{{\ttfamily 0708.2443}}].

\bibitem{Kuhn:1998uy}
J.H.~Kuhn, A.A.~Penin and A.A.~Pivovarov, \emph{{Coulomb resummation for b
  anti-b system near threshold and precision determination of alpha(s) and
  m(b)}}, \href{https://doi.org/10.1016/S0550-3213(98)00607-5}{\emph{Nucl.
  Phys. B} {\bfseries 534} (1998) 356}
  [\href{https://arxiv.org/abs/hep-ph/9801356}{{\ttfamily hep-ph/9801356}}].

\bibitem{Mannel:2021ubk}
T.~Mannel, D.~Moreno and A.A.~Pivovarov, \emph{{Master integrals for inclusive
  weak decays of heavy flavors at next-to-leading order}},
  \href{https://doi.org/10.1103/PhysRevD.104.114035}{\emph{Phys. Rev. D}
  {\bfseries 104} (2021) 114035}
  [\href{https://arxiv.org/abs/2104.13080}{{\ttfamily 2104.13080}}].

\end{thebibliography}\endgroup

\end{document}